\documentclass[12pt]{article}
\usepackage{amsmath,amsfonts,amssymb,cite}
\newcommand{\be}{\begin{equation}}
\newcommand{\ee}{\end{equation}}
\newcommand{\ba}{\begin{eqnarray}}
\newcommand{\ea}{\end{eqnarray}}

\newcommand{\la}[1]{\label{#1}}

\def\gl#1{(\ref{#1})}

\date{}
\begin{document}
\title{Spectral Design for Matrix Hamiltonians:\\ Different Methods of Constructing of\\ a Matrix Intertwining Operator}
\author{A.V. Sokolov\footnote{E-mail: avs\_avs@rambler.ru.}\\ \\{\it Deptartment of
Theoretical Physics,}\\{\it Saint-Petersburg State University,}\\{\it Ulianovskaya ul., 1, Petrodvorets,
198504 Saint-Petersburg, Russia}} 
\maketitle 
\abstract{We study intertwining relations for $n\times n$ matrix non-Hermitian, in general, one-di\-mensional Hamiltonians by $n\!\times\! n$ matrix linear differential operators with nondegenerate coefficients at $d/dx$ in the highest degree. Some methods of constructing of $n\!\times\! n$ matrix intertwining operator of the first order of general form are proposed and their interrelation is examined.  As example we construct $2\!\times\!2$ matrix Hamiltonian of general form intertwined by operator of the first order with the Hamiltonian with zero matrix potential. It is shown that one can add for the final $2\!\times\!2$ matrix Hamiltonian with respect to the initial matrix Hamiltonian with the help of intertwining operator of the first order either up to two bound states for different energy values or up to two bound states described by vector-eigenfunctions for the same energy value or up to two bound states described by vector-eigenfunction and associated vector-function for the same energy value.}

\section{Introduction}

There are two main areas of applying of matrix models with supersymmetry in Quantum Mechanics: multichannel scattering and spectral design in description of motion of spin particles in external fields. The simplest cases of such models are considered, for example, in \cite{anio88,acd88,acd90-1,anio91,aisv91,canio92,canio93,takahiro93,hgb95,basp97,lrsfv98,dc99,tr99,iknn06,fmn10} and their systematic studying is contained in \cite{acni97,gove98,ione03,cinn04,sampe03,suzko05,pepusa11,tanaka11,nk11,nika11,ni12,ka12,ka13,sokolov13,pupasov13} (see also the recent reviews \cite{anio12,pup14} and references therein). The authors of \cite{acni97} investigate intertwining of matrix Hermitian Hamiltonians by $n\times n$ first-order and $2\times 2$ second-order matrix differential operators and the corresponding supersymmetric algebras. The main result of \cite{gove98} is the formulae that provide us with the opportunity to construct for a given $n\times n$ matrix non-Hermitian, in general, Hamiltonian a new   $n\times n$ matrix Hamiltonian and an $n\times n$ matrix linear differential operator of arbitrary order with the identity matrix coefficient at $d/dx$ in the highest degree that intertwines these Hamiltonians. 

There are some shortcomings of the results of \cite{gove98}. Firstly, the formulae of \cite{gove98} are built in terms of a basis in a subspace that is invariant with respect to the initial Hamiltonian, {\it i.e.} an $n\times n$ matrix intertwining operator of the $N$-th order and the corresponding new Hamiltonian are constructed in terms of columns of $n\times nN$ matrix-valued solution ${\bf\Psi}(x)$ of the equation \begin{equation}H_+{\bf\Psi}={\bf\Psi}\Lambda,\la{gove1}\end{equation} where $H_+$ and $\Lambda$ are respectively the initial Hamiltonian and $nN\times nN$ constant matrix. It was shown in \cite{sokolov13} that one can get any $n\times n$ matrix intertwining operator of arbitrary order with arbitrary nondegenerate matrix coefficient at $d/dx$ in the highest degree and the corresponding new Hamiltonian with the help of such matrix-valued solution of (\ref{gove1}) that $\Lambda^t$ for this solution is a matrix of normal (Jordan) form. The columns of the solution ${\bf\Psi}(x)$ in this case are obviously a formal vector-eigenfunctions and formal associated vector-functions of the Hamiltonian $H_+$, where the word ``formal'' emphasizes that these vector-functions are not necessarily normalizable. It seems more easy to find formal vector-eigenfunctions and formal associated vector-functions of the Hamiltonian $H_+$ and to construct the matrix ${\bf\Psi}(x)$ from these vector-functions than to look for a matrix solution ${\bf\Psi}(x)$ of general form for \gl{gove1} as it was proposed in \cite{gove98}. Hence, the offered in \cite{sokolov13} method of constructing of a matrix intertwining operator and the corresponding new matrix Hamiltonian in terms of formal vector-eigenfunctions and associated vector-functions of the initial Hamiltonian $H_+$ (see as well the partial case of this method based on the use of formal vector-eigenfunctions only in \cite{sampe03}) allows us to simplify without the loss of generality the procedure proposed in \cite{gove98}.

Secondly, the formulae of \cite{gove98} are unnecessarily complicated since they contain {\it quasideterminants} introduced in \cite{gere}. The significantly more simple formulae in terms of usual determinants for constructing of a matrix intertwining operator with the identity matrix coefficient at $d/{dx}$ in the highest degree and the corresponding new Hamiltonian were derived in a rather sophisticated way in \cite{sampe03}. But the formulae of\cite{sampe03} were received for the partial case only where all columns of ${\bf\Psi}(x)$ are a formal vector-eigenfunctions of~$H_+$ and the intertwined Hamiltonians are Hermitian. It should be emphasized that applying of formal vector-eigenfunctions only of $H_+$ as columns in ${\bf\Psi}(x)$ results in significant narrowing of the set of received intertwining operators even in the case where $H_+$ is Hermitian. The formulae that provide us with the opportunity to build with the help of usual determinants any $n\times n$ matrix intertwining operator of arbitrary order with arbitrary constant nondegenerate matrix coefficient at $d/dx$ in the highest degree for a given $n\times n$ matrix non-Hermitian, in general, initial Hamiltonian $H_+$ and the corresponding new matrix Hamiltonian were obtained in a simple way in \cite{sokolov13}. In the partial case of \cite{sampe03} the indicated formulae of \cite{sokolov13} correspond to the formulae of \cite{sampe03}. The detailed analysis of some more shortcomings of \cite{gove98} and \cite{sampe03} can be found in \cite{sokolov13}.

The paper \cite{suzko05} contains the formulae that allow us to construct any $n\times n$ matrix differential intertwining operator of the first order with arbitrary nondegenerate matrix coefficient at $d/{dx}$ in terms of $n\times n$ matrix-valued solution ${\bf \Psi}(x)$ of the equation \gl{gove1} for the case where the Hamiltonian $H_+$ is Hermitian. As well the author of \cite{suzko05} considers the  corresponding supersymmetry algebra for the case where the mentioned coefficient at $d/{dx}$ is the identity matrix and both intertwined Hamiltonians are Hermitian, builds $n\times n$ matrix differential intertwining operators of higher orders from chains of first-order $n\times n$ matrix differential intertwining operators and investigates in details $n\times n$ matrix intertwining operators of the second order, obtained in this way.

The generalization of results of the paper \cite{sampe03} to the case of a degenerate matrix coefficient of an intertwining operator at $d/{dx}$ in the highest degree is considered in \cite{pepusa11}. The author of \cite{pupasov13} builds $n\times n$ matrix differential intertwining operators of the second order for Hermitian matrix Hamiltonians with all real-valued elements in their potentials in terms of two $n\times n$ matrix-valued solutions ${\bf\Psi}_1(x)$ and ${\bf\Psi}_2(x)$ of the equation \gl{gove1} for the matrices $\Lambda_1$ and $\Lambda_2$ respectively in its right-hand side of the form 
\[\Lambda_1=E_1I_n,\quad\Lambda_2=E_2I_n, \qquad E_1,E_2\in{\Bbb C},\] 
where $I_n$ is the identity matrix of the $n$-th order. As well the corresponding polynomial supersymmetry algebra of the second order is constructed and different applications of the obtained results are examined in \cite{pupasov13}.

The author of \cite{tanaka11} proposes to study a supersymmetry generated by two $n\times n$ matrix non-Hermitian, in general, Hamiltonians $H_+$ and $H_-$ and two $n\times n$ matrix differential operators $Q_N^+$ and $Q_N^-$ of the same order $N$ with constant coefficients proportional to the identity matrix at $({d/{dx}})^N$ that intertwine $H_+$ and $H_-$ in the opposite directions and such that the products $Q_N^+Q_N^-$ and $Q_N^-Q_N^+$ are the same polynomials with matrix coefficients of $H_+$ and $H_-$ respectively. Moreover, the operators $Q_N^+$ and $Q_N^-$ are supposed to be related one to another by some unnatural operation which is not, in general, neither transposition nor Hermitian conjugation. Hence, intertwining of $H_+$ and $H_-$ by one of the operators $Q_N^+$ and $Q_N^-$ does not lead, in general, to the intertwining of $H_+$ and $H_-$ by another of the operators $Q_N^+$ and $Q_N^-$ even if both Hamiltonians $H_+$ and $H_-$ are symmetric with respect to transposition or Hermitian. Thus, the intertwining operators $Q_N^+$ and $Q_N^-$ generate independent, in general, restrictions on the system in question. In addition, there are no in \cite{tanaka11} neither proof of existence of the considered system for arbitrary $n$ and $N$ nor any general method of constructing of this system. Only for the case $n=N=2$ the author finds general form of $H_+$, $H_-$, $Q_N^+$ and $Q_N^-$ under additional assumption that $H_+$, $H_-$ and all coefficients of the operators $Q_N^+$ and $Q_N^-$ are Hermitian.

The paper \cite{sokolov13} in addition to the formulae for constructing of arbitrary matrix intertwining operator and the corresponding new matrix Hamiltonian (see above) contains the results on existence for arbitrary $n\times n$ matrix intertwining operator of the order $N$ with arbitrary nondegenerate matrix coefficient at $(d/dx)^N$ an $n\times n$ matrix differential operator of different, in general, order $N'$ that intertwines the same Hamiltonians in the opposite direction and on the corresponding polynomial supersymmetry algebra. Earlier the case of two scalar differential operators of different, in general, orders that intertwine two scalar differential operators of partial form in the opposite directions was considered in \cite{shabat12}.  As well there are in \cite{sokolov13} the criteria of minimizability \cite{anso03,ancaso07} and of reducibility \cite{acdi95,samsonov99,anca04,anso06,sokolov07,sokolov10} of a matrix intertwining operator.

Some supersymmetric matrix models with shape invariance are investigated in \cite{nk11,nika11,ni12,ka12,ka13}. Most of the mentioned above papers on the matrix case is devoted in fact to the case of one spatial variable. The cases of two and three spacial variables are considered in \cite{anio88,anio91,aisv91,nika11,ione03,cinn04}.

The purpose of this paper is (i) to derive some methods for constructing of arbitrary $n\times n$ matrix first-order intertwining operator with arbitrary constant nondegenerate matrix coefficient at $d/dx$ and the corresponding new matrix Hamiltonian in the case where both intertwined Hamiltonians are, in general, non-Hermitian, (ii) to investigate interrelations of these methods and (iii) to demonstrate the capabilities of these methods for spectral design of matrix Hamiltonians. The present paper is organized as follows. Section~2 contains basic definitions and notation. Section 3 is devoted to derivation of some methods for constructing of any $n\times n$ matrix first-order intertwining operator with arbitrary constant nondegenerate matrix coefficient at $d/{dx}$ and of the corresponding new matrix Hamiltonian. Namely, we present the method of matrix superpotential and one more method in Subsection 3.1, the method of transformation vector-functions in Subsection 3.2 and the method of transformation matrix in Subsection 3.3. As well we examine in Section 3 the interrelations of these methods. Section~4 includes brief description of generalization of the method of transformation vector-functions to the case of matrix intertwining operator of arbitrary order. In Section 5 we present three examples that demonstrate capabilities of the methods of Section 3 for spectral design of matrix Hamiltonians. It is shown that one can add for the final $2\times 2$ matrix Hamiltonian with respect to initial $2\times 2$ matrix Hamiltonian with the help of $2\times 2$ first-order matrix intertwining operator either up to two bound states for different energy values (Subsection~5.1) or up to two bound states described by vector-eigenfunctions for the same energy value (Subsection 5.2) or up to two bound states described by vector-eigenfunction and associated vector-function for the same energy value (Subsection 5.3). In Conclusions we itemize some problems which can be considered in future papers.

\section{Basic definitions and notation}

\subsection{Intertwining relation\la{s2.1}}

Let's consider two defined on the entire axis matrix Hamiltonians of Schr\"odinger form

\kern-1mm

\[H_+=-I_n\partial^2+V_+(x),\quad H_-=-I_n\partial^2+V_-(x),\qquad
\partial\equiv{d\over{dx}},\la{h+h-2.1}\] 
where $I_n$ is the identity matrix of the $n$-th order, $n\in{\Bbb N}$, and $V_+(x)$ and $V_-(x)$ are square $n\times n$ matrices, all elements of which are sufficiently smooth and, in general, complex-valued functions. These Hamiltonians are supposed to be
{\it intertwined} by a matrix linear differential operator $Q_N^-$, so that

\kern-3mm

\begin{equation} Q_N^-H_+=H_-Q^-_N,\qquad
Q_N^-=\sum\limits_{j=0}^NX^-_j(x)\partial^j,\la{splet}\end{equation}

\kern-1mm

\noindent where $X^-_j(x)$, $j=0$, \dots, $N$ are as well square $n\times n$ matrices, all elements of which are sufficiently smooth and, in general, complex-valued functions. The operator $Q_N^-$ in this case is called {\it intertwining operator}.

It follows from \gl{splet} (see \cite{sokolov13}) that 

\kern-1mm

\[X_N^-= {\rm Const}\]

\kern-3mm

\noindent and

\kern-3mm

\begin{equation}X^-_NV_+(x)=-2X^{-\,\prime}_{N-1}(x)+V_-(x)X^-_N.\la{v12}\end{equation} 
We shall suppose below that $\det X^-_N\ne0$. In this case one can find from (\ref{v12}) the matrix potential $V_-(x)$ in terms of $V_+(x)$ and $X^-_{N-1}(x)$,

\kern-1mm

\begin{equation}V_-(x)=X^-_NV_+(x)(X^-_N)^{-1}+2X^{-\,\prime}_{N-1}(x)(X^-_N)^{-1}.\la{vmp2.1}\end{equation}

\subsection{Structure of intertwining operator kernel and transformation vector-functions \la{sjio}}

In view of \gl{splet} the kernel of the intertwining operator $Q_N^-$ is an invariant
subspace for the Hamiltonian $H_+$: 
\[H_+\ker Q_N^-\subset\ker Q_N^-.\] 
Therefore, for any basis $\Phi^-_1(x)$, \dots, $\Phi^-_d(x)$ in the kernel of $Q_N^-$, $d=\dim\ker Q_N^-=nN$ there exists a constant
square $d\times d$ matrix ${\bf T}^+\equiv\|T^+_{ij}\|$ such that
\begin{equation}H_+\Phi^-_i=\sum_{j=1}^dT^+_{ij}\Phi^-_j,\qquad i=1,\ldots,d.
\la{tm}\end{equation} 
Let us note that the Wronskian of all elements of any basis in $\ker Q_N^-$ does not vanish on the entire axis.

One can construct from the elements of the basis $\Phi^-_1(x)$, \dots, $\Phi^-_d(x)$
as from columns the $n\times d$ matrix-valued solution 
\[{\bf\Psi}(x)=\big(\Phi^-_1(x),\ldots,\Phi^-_d(x)\big)\] of the equation \gl{gove1}
and the matrix $\Lambda$ from \gl{gove1} is interrelated with the
matrix~${\bf T}^+$ by the evident equality \[\Lambda=({\bf T}^+)^t.\]

In the what follows, the {\it matrix} $\bf T$ of an intertwining operator is defined as a matrix which is constructed for the operator in the same way as the matrix ${\bf T}^+$ is constructed for $Q_N^-$. In this case, we do not specify the basis in the kernel of the intertwining operator in which the matrix $\bf T$ is chosen if we concern only spectral characteristics of the matrix, or, what is the same, spectral characteristics of the restriction of the corresponding Hamiltonian to the kernel of the considered intertwining operator (cf. with \gl{tm}).

A basis in the kernel of an intertwining operator in which the
matrix $\bf T$ of this operator has a normal (Jordan) form is called a {\it
canonical basis}. Elements of a canonical basis are called {\it
transformation vector-functions}.

If a Jordan form of the matrix $\bf T$ of an intertwining operator contains
block(s) of order higher than one, then there are in the corresponding canonical
basis not only formal vector-eigenfunction(s) of the corresponding Hamiltonian but also its formal associated vector-function(s) which are
defined as follows (see \cite{naim}).

A vector-function $\Phi_{m,i}(x)$ is called a {\it formal associated vector-function of $i$-th order} of an $n\times n$ matrix Hamiltonian
$H=-I_n\partial^2+V(x)$ for a spectral value $\lambda_m$ if
\[(H-\lambda_mI_n)^{i+1}\Phi_{m,i}\equiv 0\quad\text{and}\quad(H-\lambda_mI_n)^{i}\Phi_{m,i}\not\equiv0,\] 
where the term ``formal'' emphasizes that the vector-function $\Phi_{m,i}(x)$ is
not necessarily normalizable (not necessarily belongs to $L^2({\Bbb R},{\Bbb C}^n)$). In particular, a formal associated
vector-function of zero order $\Phi_{m,0}(x)$ is a formal
vector-eigenfunction of $H$.

A finite or infinite set of vector-functions $\Phi_{m,i}(x)$, $i=0$, 1, 2, \dots\, is called a {\it chain} of formal associated vector-functions of an $n\times n$ matrix Hamiltonian $H=-I_n\partial^2+V(x)$ for a spectral value $\lambda_m$ if \[H\Phi_{m,0}=\lambda_m\Phi_{m,0},\quad\Phi_{m,0}(x)\not\equiv0,
\qquad (H-\lambda_mI_n)\Phi_{m,i}=\Phi_{m,i-1},\quad i=1,2,3,\ldots\,.\] It is evident that $\Phi_{m,i}(x)$ in this case is a formal associated vector-function of $i$-th order of the Hamiltonian $H$ for the spectral value $\lambda_m$, $i=0$, 1, 2, \dots\,.

A chain $\Psi_{m,l}^-(x)$, $l=0$, 1, 2, \dots\, of formal associated vector-functions of the Hamiltonian $H_+$ for a spectral value $\lambda_m$ in view of the equalities 
\[(H_--\lambda_mI_n)Q_N^-\Psi_{m,l}^-=Q_N^-(H_+-\lambda_mI_n)\Psi_{m,l}^-=
Q_N^-\Psi_{m,l-1}^-,\] 
\begin{equation}l=0,1,2,\ldots,\qquad \Psi_{m,-1}^-(x)\equiv0,\la{map}\end{equation} 
that take place due to \gl{splet}, is mapped by $Q_N^-$ into a chain of formal associated vector-functions of the Hamiltonian $H_-$ for the same spectral value $\lambda_m$ with possible exception of some number of vector-functions $Q_N^-\Psi_{m,l}^-$ with lower numbers which can be identical zeroes. It is clear in view of \gl{map} that if $Q_N^-\Psi_{m,l_0}^-\equiv0$ for some $l_0$ then $Q_N^-\Psi_{m,l}^-\equiv0$ for any $l<l_0$ and if $Q_N^-\Psi_{m,l_0}^-\not\equiv0$ for some $l_0$ then $Q_N^-\Psi_{m,l}^-\not\equiv0$ for any $l>l_0$. Thus, if $l_0$ is a minimal number such that $Q_N^-\Psi_{m,l_0}^-\not\equiv0$ then one can represent the arising chain of formal associated vector-functions  of $H_-$ in the form \[\Psi_{m,l}^+(x)=Q_N^-\Psi_{m,l+l_0}^-(x), \qquad l=0,1,2,\ldots\,.\] 

\section{Methods of constructing of a first-order matrix intertwining operator\la{SO1O}}

\subsection{Method of matrix superpotential  and one more method}

Let us consider the case where two $n\times n$ matrix Hamiltonians
$H_+$ and  $H_-$ are intertwined by a first-order $n\times n$ matrix differential operator
\[Q_1^-=X_1^-\partial+X_0^-(x),\]
so that \begin{equation}Q_1^-H_+=H_-Q_1^-.\la{splet2n}\end{equation}
In view of Section \ref{s2.1} the matrix coefficient $X_1^-$ is a constant nondegenerate matrix. Thus, we can rewrite the equality \gl{splet2n} with the help of
multiplying it from the left by $(X_1^-)^{-1}$ in the form
\[\big((X_1^-)^{-1}Q_1^-\big)H_+=
\big((X_1^-)^{-1}H_-X_1^-\big)\big((X_1^-)^{-1}Q_1^-\big).\] It
follows from the latter equality that two $n\times n$ matrix
Hamiltonians \[H_+=-I_n\partial^2+V_+(x),\qquad \tilde
H_-=-I_n\partial^2+\tilde V_-(x),\quad \tilde
V_-(x)=(X_1^-)^{-1}V_-(x)X_1^-\] are intertwined by the first-order $n\times n$ matrix
differential operator
\[\tilde Q_1^-=I_n\partial+\tilde X_0^-(x),\quad\tilde X_0^-(x)=(X_1^-)^{-1}X_0^-(x),\]
so that
\begin{equation}\tilde Q_1^-H_+=\tilde H_-\tilde Q_1^-.\la{splet3n}\end{equation}

Now we shall look for general solution of the intertwining relation
\gl{splet3n}. This solution can be found (see below) in the form of
parametrization of the potentials $V_+(x)$ and $\tilde V_-(x)$ and of the superpotential $\tilde X_0^-(x)$
by $n^2$ arbitrary scalar functions which are, in general, complex-valued. After receiving of this solution general solution of
intertwining relation~\gl{splet2n} can be restored with the help of the following evident relations:
\begin{equation}V_+(x)=V_+(x),\qquad V_-(x)=X_1^-\tilde
V_-(x)(X_1^-)^{-1}, \qquad X^-_0(x)=X_1^-\tilde
X_0^-(x)\la{ge1n}\end{equation} with arbitrary nondegenerate
$n\times n$ matrix $X_1^-$.

Intertwining relation \gl{splet3n} is equivalent to two equations,
\begin{eqnarray}V_+(x)&=&-2\tilde X_0^{-\,\prime}(x)+\tilde V_-(x),\nonumber\\
V'_+(x)+\tilde X_0^-(x)V_+(x)&=&-\tilde X_0^{-\,\prime\prime}(x)+
\tilde V_-(x)\tilde X_0^-(x).\la{eq5n}\end{eqnarray} It follows from
the first of these equations that $V_+(x)$ and $\tilde V_-(x)$ can
be represented in the form \begin{equation}V_+(x)=V_0(x)-\tilde
X_0^{-\,\prime}(x),\qquad\tilde V_-(x)=V_0(x)+\tilde
X_0^{-\,\prime}(x),\la{vpmn}\end{equation} with some unknown
$n\times n$ matrix-valued function $V_0(x)$. This function by virtue of the
second equation in \gl{eq5n} satisfies the equation
\[V'_0(x)=[V_0(x),\tilde X_0^-(x)]+\tilde X_0^{-\,\prime}(x)\tilde
X_0^-(x)+\tilde X_0^-(x)\tilde X_0^{-\,\prime}(x).\] The latter
equation after the change \begin{equation}V_0(x)=U_0(x)+\big(\tilde
X_0^-(x)\big)^2\la{changen}\end{equation} transforms into
\begin{equation}U'_0(x)=[U_0(x),\tilde X_0^-(x)],\la{u'n}\end{equation} where $U_0(x)$ is new unknown $n\times n$ matrix-valued function.

General solution of the equation \gl{u'n} can be constructed in some ways. One of these ways is the following. One can consider all $n^2$ elements of the matrix superpotential $\tilde X_0^-(x)$ as arbitrary complex-valued, in general, parametrizing functions. Then the equation \gl{u'n} is a system of~$n^2$ linear first-order ordinary differential equations with respect to elements of the matrix $U_0(x)$. General solution of this system is parametrized by $n^2$ arbitrary functions (elements of $\tilde X_0^-(x)$) and $n^2$ arbitrary complex, in general, constants.

Another way to find general solution of the system \gl{u'n} is to take all $n^2$ elements of the matrix $U_0(x)$ as arbitrary complex-valued, in general, parametrizing functions. Then the equation \gl{u'n} is a system of $n^2$ linear algebraic equations (SLAE) with respect to elements of the matrix superpotential $\tilde X_0^-(x)$. This SLAE is degenerate, in general, and conditions of its compatibility lead to restrictions on elements of the matrix $U_0(x)$ and, consequently, to a decrease in the number of independent functions among the elements of the matrix $U_0(x)$. Nevertheless, the total number of independent  parametrizing functions is again equal to $n^2$ due to the appearance of free variables and to the evident fact that the number of compatibility conditions is equal to the number of appearing free variables. Thereby, general solution of SLAE \gl{u'n} is parametrized by $n^2$ arbitrary functions (independent elements of $U_0(x)$ and free variables).

The latter of two described above ways of parametrization of general solution of \gl{u'n} is more suitable than the former since the latter way in contrast to the former leads to explicit parametrizing formulae. Two more ways to construct general solution of \gl{u'n} will be presented in the following two subsections.

Thus, general solution of intertwining relation \gl{splet3n} is given
in view of \gl{vpmn} and \gl{changen} by the formulae
\begin{equation}V_+(x)=U_0(x)+(\tilde X_0^-(x))^2-\tilde
X_0^{-\,\prime}(x),\qquad \tilde V_-(x)=U_0(x)+(\tilde
X_0^-(x))^2+\tilde X_0^{-\,\prime}(x),\la{ge2n}\end{equation} 
where $U_0(x)$ and $\hat X_0^-(x)$ are found in one of the described ways. Hence, general solution of intertwining relation
\gl{splet2n} is given by \gl{ge1n} together with \gl{ge2n}.

It is evident that in view of \gl{ge2n} the Hamiltonians $H_+$ and
$\tilde H_-$ can be represented in the form
\begin{equation}H_+=\tilde Q_1^+\tilde Q_1^-+U_0(x),\quad\tilde
H_-=\tilde Q_1^-\tilde Q_1^++U_0(x),\qquad \tilde
Q_1^+=-I_n\partial+\tilde X_0^-(x).\la{fakt1n}\end{equation} Moreover,
the intertwining relation \gl{splet3n} for these Hamiltonians is
provided by the condition
\begin{equation}[U_0(x),\tilde Q_1^-]=0\la{com1n}\end{equation} which
is equivalent to the equation \gl{u'n}.

Intertwining of the Hamiltonians $H_+$ and $\tilde H_-$ by the
operator $\tilde Q_1^+$, \begin{equation}H_+\tilde Q_1^+=\tilde Q_1^+\tilde H_-\la{q1+int19}\end{equation}
is equivalent (in the case if this intertwining takes place) to the condition
\begin{equation}[U_0(x),\tilde Q_1^+]=0.\la{f22''n}\end{equation}
The latter condition is equivalent, in
turn, in view of \gl{com1n} to the equality
\begin{equation}[U_0(x),\partial]=0,\la{f22'''n}\end{equation} {\it i.e.} to
independence of all elements of the matrix $U_0(x)$ from $x$.

By virtue of \gl{ge1n} and \gl{fakt1n} general solution of intertwining relation \gl{splet2n} can be represented in the form
\begin{equation}H_+=Q_1^+Q_1^-+U_0(x),\quad H_-=Q_1^-Q_1^++U(x),\qquad
U(x)=X_1^-U_0(x)(X_1^-)^{-1},\la{f22'n}\end{equation}
\begin{equation}Q_1^-\equiv X_1^-\partial+X_0^-(x)=X_1^-\tilde Q_1^-,\qquad
X_0^-(x)=X_1^-\tilde X_0^-(x),\la{f22n}\end{equation}
\begin{equation}Q_1^+\equiv X_1^+\partial+X_0^+(x)=\tilde Q_1^+(X_1^-)^{-1},\qquad
X_1^+=-(X_1^-)^{-1},\quad X_0^+(x)=\tilde
X_0^-(x)(X_1^-)^{-1}.\la{f26n}\end{equation} Intertwining \gl{splet2n}
for the constructed Hamiltonians $H_+$ and
$H_-$ is valid due to the relation
\[Q_1^-U_0(x)=U(x)Q_1^-\] which follows from \gl{com1n}, \gl{f22'n}
and \gl{f22n}. It is easy to see that intertwining
\[H_+Q_1^+=Q_1^+H_-\] is equivalent to the relation \[U_0(x)Q_1^+=Q_1^+U(x)\]
which is equivalent, in turn, by \gl{f22'n} and \gl{f26n} to
\gl{f22''n} and, consequently, to \gl{f22'''n}. The latter is
obviously equivalent to independence of $U_0(x)$ and $U(x)$ from
$x$.

\subsection{Method of transformation vector-functions \la{IOnTF}}

Let us consider $H_+$ as known initial $n\times n$ matrix Hamiltonian and $\Phi^-_l(x)$, $l=1$, \dots, $n$ be a set of formal associated vector-functions of
$H_+$ such that \begin{equation}H_+\Phi_l=\lambda_l\Phi_l+\sigma_l\Phi_{l+1},
\qquad \sigma_l=\begin{cases}1,&\text{$\Phi^-_l(x)$ is not a formal vector-eigenfunction},\\
0,& \text{$\Phi^-_l(x)$ is a formal vector-eigenfunction,}\end{cases}\la{hp3.2}\end{equation} \begin{equation}\Phi_l^-(x)\equiv\begin{pmatrix}\varphi^-_{l1}(x)\\\varphi^-_{l2}(x)\\\vdots\\\varphi^-_{ln}(x)\end{pmatrix},\quad 
l=1,\ldots,n,\qquad\Phi^-_{n+1}(x)\equiv\begin{pmatrix}0\\0\\\vdots\\0\end{pmatrix},\la{phl3.2}\end{equation} 
where $\lambda_l$ is the spectral value of $H_+$ corresponding to $\Phi^-_l(x)$, $l=1$,
\dots, $n$, and \begin{equation}\lambda_{l+1}=\lambda_l\quad\text{if} \quad \sigma_l=1,\qquad
l=1,\ldots, n-1.\la{lal3.2}\end{equation} 
We shall suppose that the Wronskian of these vector-functions 
\begin{equation} W(x)\equiv\begin{vmatrix}\varphi^-_{11}(x)&\varphi^-_{12}(x)&\dots&\varphi^-_{1n}(x)\\
\varphi^-_{21}(x)&\varphi^-_{22}(x)&\dots&\varphi^-_{2n}(x)\\
\vdots&\vdots&\ddots&\vdots\\
\varphi^-_{n1}(x)&\varphi^-_{n2}(x)&\dots&\varphi^-_{nn}(x)
\end{vmatrix}\la{wron3.2}\end{equation} does not vanish on the entire axis. In this case we can consider all $n^2$ elements of the matrix potential $V_+(x)$ of the Hamiltonian $H_+$ as functions that implicitly parametrize vector-functions $\Phi_l^-(x)$, $l=1$, \dots, $n$ and, consequently, the $n\times n$ matrix superpotential $X_0^-(x)$ from the intertwining operator $Q_1^-=X_1^-\partial+X_0^-(x)$ and the $n\times n$ matrix potential $V_-(x)$ of the final Hamiltonian $H_-$ which will be constructed below in terms of $\Phi_l^-(x)$, $l=1$, \dots, $n$.

It is possible as well to suppose that the Hamiltonian $H_+$ is not known initially and that the vector-functions \gl{phl3.2} are arbitrary vector-functions with complex-valued, in general, components such that the Wronskian \gl{wron3.2} does not vanish on the entire axis. In this case one can choose arbitrarily constants $\lambda_l\in{\Bbb C}$ and $\sigma_l\in\{0,1\}$, $l=1$, \dots, $n$, so that the conditions \gl{lal3.2} are valid, and thereafter to find the only $n\times n$ matrix potential 
\[V_+(x)\equiv\|v^+_{ij}(x)\|\] 
of the Hamiltonian $H_+$ such that the relations \gl{hp3.2} hold with the help of solving of the following SLAEs:
\[\begin{cases}\begin{matrix}v^+_{l1}\varphi^-_{11}+v^+_{l2}\varphi^-_{12}+\dots+v^+_{ln}\varphi^-_{1n}&=&
\varphi^{-\prime\prime}_{1l}+\lambda_1\varphi^-_{1l}+\sigma_1\varphi^-_{2l},\hfill\\
v^+_{l1}\varphi^-_{21}+v^+_{l2}\varphi^-_{22}+\dots+v^+_{ln}\varphi^-_{2n}&=&
\varphi^{-\prime\prime}_{2l}+\lambda_2\varphi^-_{2l}+\sigma_2\varphi^-_{3l},\hfill\\
\vdots&&\\
v^+_{l1}\varphi^-_{n1}+v^+_{l2}\varphi^-_{n2}+\dots+v^+_{ln}\varphi^-_{nn}&=&
\varphi^{-\prime\prime}_{nl}+\lambda_n\varphi^-_{nl}+\sigma_n\varphi^-_{n+1,l},\end{matrix}
\end{cases}\qquad l=1,\ldots, n,\] which are equivalent to \gl{hp3.2}. Any of these SLAEs possesses by the only solution due to the fact that $W(x)$ does not vanish on the entire axis and, thus, elements of $V_+(x)$ can be found with the help of Cramer formulae:
\begin{equation}v^+_{lj}\!=\!{1\over{W}}\!\begin{vmatrix}\varphi^-_{11}&\ldots&\varphi^-_{1,j-1}&
\varphi^{-\prime\prime}_{1l}\!+\!\lambda_1\varphi^-_{1l}\!+\!\sigma_1\varphi^-_{2l}&\varphi^-_{1,j+1}&
\ldots&\varphi^-_{1n}\\
\varphi^-_{21}&\ldots&\varphi^-_{2,j-1}&
\varphi^{-\prime\prime}_{2l}\!+\!\lambda_2\varphi^-_{2l}\!+\!\sigma_2\varphi^-_{3l}&\varphi^-_{2,j+1}&
\ldots&\varphi^-_{2n}\\
\vdots&\ddots&\vdots&\vdots&\vdots&\ddots&\vdots\\
\varphi^-_{n1}&\ldots&\varphi^-_{n,j-1}&
\!\!\varphi^{-\prime\prime}_{nl}\!+\!\lambda_n\varphi^-_{nl}\!+\!\sigma_n\varphi^-_{n+1,l}\!\!&\varphi^-_{n,j+1}&
\ldots&\varphi^-_{nn}\end{vmatrix}\!,\,\,\,\,\,\, l,j\!=\!1,\,\ldots,\,n.\la{vlj3.2}\end{equation} 
In this case one can consider all $n^2$ components of $\Phi_l^-(x)$, $l=1$, \dots, $n$ as parametrizing functions. Then
elements of $V_+(x)$ are parametrized by these components explicitly with the help of \gl{vlj3.2} and explicit parametrizations of 
$X^-_0(x)$ and $V_-(x)$ in terms of considered components will be presented below. Thus, the parametrization in terms of components of $\Phi_l^-(x)$, $l=1$, \dots, $n$ is more suitable than the parametrization in terms of elements of $V_+(x)$ since the former is explicit and the latter is implicit.

Let us now construct an auxiliary operator $\tilde Q_1^-$, operators $Q_1^-$ and $Q_1^+$ and Hamiltonian $H_-$ and thereafter check that $H_+$ and $H_-$ are intertwined by $Q_1^-$.

There is the only $n\times n$ matrix linear differential operator $\tilde Q_1^-$ of the form
\[\tilde Q_1^-\equiv I_n\partial+\tilde X_0^-(x),\la{q1m3.2}\] 
kernel of which contains all vector-functions \gl{phl3.2}. This operator can be found with the help of the following evident explicit formula, \[\tilde Q_1^-={1\over{W(x)}}\begin{vmatrix}
\varphi^-_{11}(x)&\varphi^-_{12}(x)&\ldots&\varphi^-_{1n}(x)&\Phi'_1(x)\\
\varphi^-_{21}(x)&\varphi^-_{22}(x)&\ldots&\varphi^-_{2n}(x)&\Phi'_2(x)\\
\vdots&\vdots&\ddots&\vdots&\vdots\\
\varphi^-_{n1}(x)&\varphi^-_{n2}(x)&\ldots&\varphi^-_{nn}(x)&\Phi'_n(x)\\
P_1&P_2&\ldots&P_{n}&I_n\partial
\end{vmatrix},\hfill\] \begin{equation}P_l\,\Phi=\varphi_l,\qquad\forall\,\,\Phi(x)\equiv\begin{pmatrix}
\varphi_1(x)\\\varphi_2(x)\\\vdots\\\varphi_n(x).\end{pmatrix},\qquad l=1,\ldots,n,\la{qtm3.2}\end{equation} 
where during calculation of the determinant in each of its terms the corresponding of the operators $P_1$, \dots,  $P_n$, $I_n\partial$ must be placed on the last position. It is not hard to see in view of \gl{qtm3.2} that $l$-th column of the matrix $\tilde X_0^-(x)$ is equal to 
\begin{equation}-{1\over{W(x)}}\begin{vmatrix}
\varphi^-_{11}(x)&\ldots&\varphi^-_{1,l-1}(x)&\Phi'_1(x)&\varphi^-_{1,l+1}(x)&\ldots&\varphi^-_{1n}(x)\\
\varphi^-_{21}(x)&\ldots&\varphi^-_{2,l-1}(x)&\Phi'_2(x)&\varphi^-_{2,l+1}(x)&\ldots&\varphi^-_{2n}(x)\\
\vdots&\ddots&\vdots&\vdots&\vdots&\ddots&\vdots\\
\varphi^-_{n1}(x)&\ldots&\varphi^-_{n,l-1}(x)&\Phi'_n(x)&\varphi^-_{n,l+1}(x)&\ldots&\varphi^-_{nn}(x)\\
\end{vmatrix},\qquad l\!=\!1,\ldots,n.\la{Xl3.2}\end{equation}
Using the operator $\tilde Q_1^-$ and arbitrary nondegenerate matrix $X_1^-$ one can construct the operators $Q_1^-$ and $Q_1^+$ with the help of the formulae \gl{f22n} and \gl{f26n} with $\tilde Q_1^+=-I_n\partial+\tilde X_0^-(x)$,
represent the Hamiltonian $H_+$ in the form 
\begin{equation}H_+=Q_1^+Q_1^-+U_0(x),\qquad U_0(x)=V_+(x)-(\tilde X_0^-(x))^2+
\tilde X_0^{-\prime}(x)\la{h+3.2}\end{equation} 
(cf. with (\ref{ge2n}) and (\ref{f22'n})) and build new Hamiltonian of Schr\"odinger form
\[H_-\equiv-I_n\partial^2+V_-(x)=Q_1^-Q_1^++U(x),\qquad U(x)=X_1^-U_0(x)(X_1^-)^{-1},\]
\begin{eqnarray}V_-(x)&=&X_1^-[(\tilde X_0^{-}(x))^2+\tilde X_0^{-\prime}(x)+U_0(x)](X_1^-)^{-1}=
X_1^-[V_+(x)+2\tilde X_0^{-\prime}(x)](X_1^-)^{-1}\nonumber\\
&=&X_1^-V_+(x)(X_1^-)^{-1}+2X_0^{-\prime}(x)(X_1^-)^{-1}\la{h-3.2}\end{eqnarray}
(cf. with \gl{vmp2.1} and \gl{f22'n}).

We shall check now that the Hamiltonians $H_+$ and $H_-$ are intertwined by $Q_1^-$ in ac\-cor\-dan\-ce with \gl{splet2n}. This intertwining in view of \gl{h+3.2} and \gl{h-3.2} is equivalent to the condition 
\begin{equation}Q_1^-U_0(x)-U(x)Q_1^-=0.\la{quuq3.2}\end{equation} 
The left-hand part of \gl{quuq3.2} by virtue of \gl{f22n}, \gl{h+3.2} and \gl{h-3.2} is an $n\!\times\! n$ matrix-valued function and the following chain is valid due to the construction of $Q_1^-$ and to \gl{hp3.2} and \gl{h+3.2},
\begin{eqnarray}[Q_1^-U_0(x)-U(x)Q_1^-]\Phi_l
&=&[Q_1^-H_+-Q_1^-Q_1^+Q_1^--U(x)Q_1^-]\Phi_l\nonumber\\
&=&Q_1^-H_+\Phi_l=Q_1^-[\lambda_l\Phi_l+\sigma_l\Phi_{l+1}]\nonumber\\
&=&0,\qquad l=1,\ldots,n.\nonumber\end{eqnarray} 
Thus, in view of the fact that the Wronskian $W(x)$ of vector-functions $\Phi_l(x)$, $l=1$, \dots, $n$ does not vanish on the entire axis we have that the condition \gl{quuq3.2} takes place and, consequently, the operator $Q_1^-$ intertwines the Hamiltonians $H_+$ and $H_-$.

Let us note that the condition that the Wronskian $W(x)$ does not vanish on the entire axis provides existence and smoothness (absence of pole(s)) for all considered in this subsection matrix-valued functions $V_+(x)$, $V_-(x)$, $\tilde X_0^-(x)$, $U_0(x)$ and $U(x)$ and as well for the coefficients $X_0^-(x)$ and $X_0^+(x)$ of the operators $Q_1^-$ and $Q_1^+$ (see \gl{f22n} and \gl{f26n}). 

All objects of this subsection coincide with the denoted in the same way objects of the previous subsection if to choose vector-functions $\Phi_l^-(x)$, $l=1$, \dots, $n$ in this subsection as elements of a canonical basis in the kernel of the intertwining operator $Q_1^-$ from the previous subsection. This statement is valid in view of the fact that matrix linear first-order differential operator with fixed nondegenerate matrix coefficient at $\partial$ is specified uniquely by a basis in its kernel. Thus, any solution of intertwining \gl{splet2n} with nondegenerate matrix coefficient $X_1^-$ can be constructed as well by the method proposed in this subsection and general solution of the equation \gl{u'n} can be presented in the form of explicit parametrization of $U_0(x)$ and $\tilde X_0^-(x)$ by $n^2$ components of vector-functions $\Phi_l^-(x)$, $l=1$, \dots, $n$ and constants $\lambda_l$ and $\sigma_l$, $l=1$, \dots, $n$ with the help of the formulae \gl{vlj3.2}, \gl{Xl3.2} and \gl{h+3.2}.

\subsection{Method of transformation vector-functions vs method of transformation matrix}

Using the transformation vector-functions $\Phi_l^-(x)$, $l=1$, \dots, $n$ of Subsection \ref{IOnTF}, one can construct the matrix 
\begin{equation}{\bf\Phi}^-(x)=\begin{pmatrix}\varphi^-_{11}(x)&\varphi^-_{21}(x)&\ldots&\varphi^-_{n1}(x)\\
\varphi^-_{12}(x)&\varphi^-_{22}(x)&\ldots&\varphi^-_{n2}(x)\\
\vdots&\vdots&\ddots&\vdots\\
\varphi^-_{1n}(x)&\varphi^-_{2n}(x)&\ldots&\varphi^-_{nn}(x)
\end{pmatrix}.\la{tm33}\end{equation} This matrix, the Hamiltonian $H_+$ and the matrix ${\bf T}_1^+$, {\it i.e} the matrix $\bf T$ of the intertwining operator $Q_1^-$ in the basis $\Phi_l^-(x)$, $l=1$, \dots, $n$ are interrelated (see Subsection \ref{sjio}) by the equality 
\begin{equation}H_+{\bf\Phi}^-={\bf \Phi}^-({\bf T}_1^+)^t,\la{h+f=33}\end{equation} which is equivalent to equalities \gl{hp3.2}.
With the help of the matrix ${\bf\Phi}^-(x)$ one can represent~\cite{gove98} the intertwining operator $\tilde Q_1^-$ in the form 
\begin{eqnarray}\tilde Q_1^-&=&I_n\partial-{\bf \Phi}^{-\prime}(x)({\bf\Phi}^-(x))^{-1}\la{qt1-33}\\
\Rightarrow\quad Q_1^-&=&X_1^-[I_n\partial-{\bf \Phi}^{-\prime}(x)({\bf\Phi}^-(x))^{-1}],\la{qm1-43}\end{eqnarray} 
where \gl{qt1-33} holds due to the following chain 
\begin{eqnarray}[I_n\partial-\!\!\!\!\!&{\bf \Phi}^{-\prime}(x)&\!\!\!\!\!({\bf\Phi}^-(x))^{-1}]{\bf\Phi}^-(x)=0\nonumber\\
&\Rightarrow&\quad
[I_n\partial-{\bf \Phi}^{-\prime}(x)({\bf\Phi}^-(x))^{-1}]\Phi_l^-(x)=0,\qquad l=1,\ldots,n\nonumber\\
&\Rightarrow&\quad
\ker[I_n\partial-{\bf \Phi}^{-\prime}(x)({\bf\Phi}^-(x))^{-1}]=\ker\tilde Q_1^-.\nonumber\end{eqnarray}
Thus, there is another formula for finding of the matrix $\tilde X_0^-(x)$:
\begin{equation}\tilde X_0^-(x)=-{\bf \Phi}^{-\prime}(x)({\bf\Phi}^-(x))^{-1}.\la{tx38}\end{equation}
The equalities \gl{qt1-33} and \gl{qm1-43} for the corresponding partial cases were found earlier in \cite{acd88,sampe03,suzko05}.

One can represent the Hamiltonians $H_+$ and $\tilde H_-$ with the help of the matrix ${\bf\Phi}^-(x)$ in the form
\begin{eqnarray}H_+&=&\tilde Q_1^+\tilde Q_1^-+{\bf\Phi}^-(x)({\bf T}_1^+)^t({\bf\Phi}^-(x))^{-1},\la{h+qq33}\\
\tilde H_-&=&\tilde Q_1^-\tilde Q_1^++{\bf\Phi}^-(x)({\bf T}_1^+)^t({\bf\Phi}^-(x))^{-1}\la{th-qq33}\\
&&\nonumber\\
\Rightarrow\qquad H_+&=& Q_1^+ Q_1^-+{\bf\Phi}^-(x)({\bf T}_1^+)^t({\bf\Phi}^-(x))^{-1},\nonumber\\
H_-&=&Q_1^- Q_1^++X_1^-{\bf\Phi}^-(x)({\bf T}_1^+)^t({\bf\Phi}^-(x))^{-1}(X_1^-)^{-1},\nonumber\end{eqnarray}
where \gl{h+qq33} and \gl{th-qq33} take place due to the equalities \gl{fakt1n}, \gl{h+f=33} and
\[[\tilde Q_1^+\tilde Q_1^-+{\bf\Phi}^-(x)({\bf T}_1^+)^t({\bf\Phi}^-(x))^{-1}]{\bf\Phi}^-(x)={\bf\Phi}^-(x)({\bf T}_1^+)^t\]
and to the facts that the Wronskian $W(x)\equiv\det {\bf \Phi}^-(x)$ does not vanish on the entire axis and the right-hand part of \gl{h+qq33} is a matrix Hamiltonian of Schr\"odinger form. The formulae \gl{h+qq33} and \gl{th-qq33} were received earlier for the corresponding partial cases in \cite{sampe03,suzko05}.

It follows from \gl{fakt1n} and \gl{h+qq33} that \begin{equation}U_0(x)={\bf\Phi}^-(x)({\bf T}_1^+)^t({\bf\Phi}^-(x))^{-1}.\la{u0rep*}\end{equation} Hence, the spectrum of the matrix $U_0(x)$ does not depend on $x$. Moreover, since the vector-functions $\Phi_l^-(x)$, $l=1$, \dots, $n$ constitute a canonical basis in $\ker Q_1^-$ and thereby the matrix ${\bf T}_1^+$ is of normal (Jordan) form, so a normal (Jordan) form of $U_0(x)$ coincides with ${\bf T}_1^+$ up to possible permutation of Jordan blocks. In the particular case where all vector-functions $\Phi_l^-(x)$,  $l=1$, \dots, $n$ are formal vector-eigenfunctions of $H_+$ for the same spectral value $\lambda_0=\lambda_1=\ldots=\lambda_n$,  the matrix $U_0(x)$ takes obviously the form \[U_0(x)=\lambda_0I_n.\]

Thus, in view of the results of Subsection \ref{IOnTF} any solution of the intertwining \gl{splet2n} with nondegenerate matrix $X_1^-$ can be constructed as well in terms of a matrix of the form~\gl{tm33} and general solution of the equation \gl{u'n} can be presented in the form of explicit parametrization of $U_0(x)$ and $\tilde X_0^-(x)$ by $n^2$ components of vector-functions $\Phi_l^-(x)$, $l=1$, \dots, $n$ and constants $\lambda_l$ and $\sigma_l$, $l=1$, \dots, $n$ with the help of the formulae \gl{tx38} and \gl{u0rep*}.

\section{Constructing of a higher-order matrix intertwi\-ning operator: method of transformation vector-functions\la{GCnn}}

It is possible to build chains of first-order matrix intertwining operators with the help of the formulae of Section \ref{SO1O} and as well higher-order matrix intertwining operators in the form of products of elements of such chains. Results of this type can be found, for example, in \cite{sampe03,suzko05} and in the more general form in Remark 1 of \cite{sokolov12}. But the indicated way of constructing of higher-order matrix intertwining operators is rather restricted since \cite{sokolov13} for any $n\geqslant2$ and $N\geqslant2$ there are $n\times n$ matrix intertwining operators of the $N$-th order that cannot be represented in the form of products of matrix intertwining operators of the lower orders. We present below the method that generalizes method of Subsection~\ref{IOnTF} and allows to construct any $n\times n$ matrix intertwining operator of arbitrary order $N$ with arbitrary constant nondegenerate matrix coefficient at $\partial^N$ and the corresponding final matrix Hamiltonian in terms of transformation vector-functions.

Let us consider $H_+$ as known initial $n\times n$ matrix Hamiltonian and $\Phi_l^-(x)$, $l=1$, \dots, $nN$, $N\in{\Bbb N}$ be a set of formal associated vector-functions of $H_+$ such that the formulae (\ref{hp3.2}) and (\ref{phl3.2}) take place for any $l=1$, \dots, $nN$, $\Phi^-_{nN+1}(x)\equiv0$, the condition (\ref{lal3.2}) is valid for any $l=1$, \dots, $nN-1$ and the Wronskian of these vector-functions
\[ W(x)\equiv\begin{vmatrix}\varphi^-_{11}&\dots&\varphi^-_{1n}& \varphi^{-\prime}_{11}&\dots&\varphi^{-\prime}_{1n}&\ldots&(\varphi^-_{11})^{(N-1)}&\dots&(\varphi^-_{1n})^{(N-1)}\\
\varphi^-_{21}&\dots&\varphi^-_{2n}&\varphi^{-\prime}_{21}&\dots&\varphi^{-\prime}_{2n}&\ldots&
(\varphi^-_{21})^{(N-1)}&\dots&(\varphi^-_{2n})^{(N-1)}\\
\vdots&\ddots&\vdots&\vdots&\ddots&\vdots&\ddots&\vdots&\ddots&\vdots\\
\varphi^-_{nN,1}\!\!&\dots&\!\!\varphi^-_{nN,n}&\varphi^{-\prime}_{nN,1}\!\!&\dots&\!\!\varphi^{-\prime}_{nN,n}\!\!&\ldots&
\!\!(\varphi^-_{nN,1})^{(N-1)}\!\!&\dots&\!\!(\varphi^-_{nN,n})^{(N-1)}
\end{vmatrix}\la{wron5}\] 
does not vanish on the entire axis. There is the only $n\times n$ matrix linear differential operator of the $N$-th order $Q_N^-$ with arbitrary nondegenerate constant $n\times n$ matrix coefficient $X_N^-$ at $\partial^N$, kernel of which contains all vector-functions $\Phi_l^-(x)$, $l=1$, \dots, $nN$, and, moreover, one can find this operator with help of the following evident formula,
\begin{equation}Q_N^-\!=\!{1\over{W}}
X_N^-\!\begin{vmatrix}\varphi^-_{11}&\dots&\varphi^-_{1n}\quad \varphi^{-\prime}_{11}&\!\!\dots\!\!&\varphi^{-\prime}_{1n}&\!\!\ldots\!\!&(\varphi^-_{11})^{(N-1)}&\!\!\dots\!\!&(\varphi^-_{1n})^{(N-1)}&(\Phi^-_1)^{(N)}\\
\varphi^-_{21}&\dots&\varphi^-_{2n}\quad
\varphi^{-\prime}_{21}&\!\!\dots\!\!&\varphi^{-\prime}_{2n}&\!\!\ldots\!\!&
(\varphi^-_{21})^{(N-1)}&\!\!\dots\!\!&(\varphi^-_{2n})^{(N-1)}&(\Phi^-_2)^{(N)}\\
\vdots&\ddots&\vdots\qquad\vdots&\!\!\ddots\!\!&\vdots&\!\!\ddots\!\!&\vdots&\!\!\ddots\!\!&\vdots&\vdots\\
\varphi^-_{nN,1}\!\!\!\!\!&\dots&\!\!\!\!\!\!\varphi^-_{nN,n}\,\,\varphi^{-\prime}_{nN,1}\!\!\!\!&\!\!\ldots\!\!&\!\!\!\!\varphi^{-\prime}_{nN,n}\!\!\!\!&\!\!\ldots\!\!&\!\!\!(\varphi^-_{nN,1})^{(N-1)}\!\!\!\!&\!\!\dots\!\!&\!\!\!(\varphi^-_{nN,n})^{(N-1)}\!\!&\!\!(\Phi^-_{nN})^{(N)}\\
P_1&\ldots&P_n\quad P_1\partial&\!\!\ldots\!\!&P_n\partial&\!\!\ldots\!\!&P_1\partial^{N-1}&\!\!\ldots\!\!&P_n\partial^{N-1}&I_n\partial^N
\end{vmatrix}\!,\la{qNrep5}\end{equation}
where $P_1$, \dots, $P_n$ are the same projection operators as in (\ref{qtm3.2}) and during calculation of the determinant \gl{qNrep5} in each of its terms the corresponding of the operators $P_1$, \dots, $P_n$, $P_1\partial$, \dots, $P_n\partial$, $P_1\partial^{N-1}$, \dots, $P_n\partial^{N-1}$, $I_n\partial^N$ must be placed on the last position. 
It fol\-lows from \gl{qNrep5} that $l$-th column of the matrix coefficient $X_j^-(x)$ of $Q_N^-$ (see (\ref{splet})) is equal~to
\[-{1\over{W}}X_N^-\,\,
\begin{matrix}\big|&\!\!\!\varphi^-_{11}&\dots&\varphi^-_{1n}&\varphi^{-\prime}_{11}&\dots&\varphi^{-\prime}_{1n}&\ldots\\
\Big|&\!\!\!\varphi^-_{21}&\dots&\varphi^-_{2n}&\varphi^{-\prime}_{21}&\dots&\varphi^{-\prime}_{2n}&\ldots\\
\bigg|&\!\!\!\vdots&\ddots&\vdots&\vdots&\ddots&\vdots&\ddots\\
\Big|&\!\!\!\varphi^-_{nN,1}&\dots&\varphi^-_{nN,n}&\varphi^{-\prime}_{nN,1}&\dots&\varphi^{-\prime}_{nN,n}&\ldots
\end{matrix}\qquad\qquad\qquad\qquad\qquad\qquad\quad\]

\[\quad\qquad \begin{matrix}(\varphi^-_{1,l-1})^{(j)}&(\Phi^-_1)^{(N)}&
(\varphi^-_{1,l+1})^{(j)}&\ldots&(\varphi^-_{11})^{(N-1)}&\dots&(\varphi^-_{1n})^{(N-1)}&\Big|\\
(\varphi^-_{2,l-1})^{(j)}&(\Phi^-_2)^{(N)}&(\varphi^-_{2,l+1})^{(j)}&\ldots&
(\varphi^-_{21})^{(N-1)}&\dots&(\varphi^-_{2n})^{(N-1)}&\Big|\\
\vdots&\vdots&\vdots&\ddots&\vdots&\ddots&\vdots&\bigg|\\
(\varphi^-_{nN,l-1})^{(j)}&(\Phi^-_{nN})^{(N)}&(\varphi^-_{nN,l+1})^{(j)}&\ldots&(\varphi^-_{nN,1})^{(N-1)}\!&\dots&(\varphi^-_{nN,n})^{(N-1)}\!\!\!&\Big|\end{matrix},\]

\begin{equation} l=1,\ldots,n,\qquad j=0,\dots,N-1.\la{tilstx5}\end{equation}
The operator $Q_N^-$ intertwines \cite{sokolov13} the initial Hamiltonian $H_+$ with some new $n\times n$ matrix Hamiltonian of Schr\"odinger form $H_-\equiv-I_n\partial^2+V_-(x)$ according to (\ref{splet}) and the potential $V_-(x)$ of $H_-$ can be found with the help of (\ref{vmp2.1}) and (\ref{tilstx5}) with $j=N-1$.

It should be emphasized that the condition that the Wronskian $W(x)$ is nonvanishing on the entire axis guarantees in view of (\ref{vmp2.1}), (\ref{qNrep5}) and (\ref{tilstx5}) existence for $Q_N^-$ and smoothness (absence of pole(s)) for the matrix-valued functions $X_0^-(x)$, \dots, $X_{N-1}^-(x)$ and $V_-(x)$.
The partial case of the representation of $Q_N^-\Phi$ for arbitrary $n$-dimensional vector-function $\Phi(x)$ with the help of (\ref{qNrep5}) and of the representation of $V_-(x)$ with the help of (\ref{vmp2.1}) and (\ref{tilstx5}) with $j=N-1$ when all vector-functions $\Phi_l^-(x)$, $l=1$, \dots, $nN$ are formal vector-eigenfunctions of the Hamiltonian $H_+$ and $X_N^-=I_n$ was found in \cite{sampe03}. 

The fact that any $n\!\times\! n$ matrix intertwining operator of arbitrary order $N$ with arbitrary nondegenerate constant matrix coefficient at $\partial^N$ can be obtained by the method presented in this section is a corollary of the facts that (i) for any operator of this type there is a canonical basis in its kernel, the Wronskian of which does not vanish on the entire axis and (ii) an $n\times n$ matrix linear differential operator of the order $N$ with a given nondegenerate constant matrix coefficient at $\partial^N$ is uniquely determined by a basis in its kernel.

\section{Examples: case $n=2$, $N=1$}

\def\ch{\,{\rm{ch}}\,}
\def\sh{\,{\rm{sh}}\,}

In this section we present some examples of constructing of $2\times2$ matrix linear differential intertwining operators of the first order $Q_1^-$ and the corresponding to them new $2\times2$ matrix Hamiltonians $H_-$ of Schr\"odinger form with the help of the methods of Section~\ref{SO1O}. As well, we demonstrate by dint of these examples the capabilities of the methods for spectral design of matrix Hamiltonians. As initial $2\times2$ matrix Hamiltonian $H_+$ we shall use the Hamiltonian of Schr\"odinger form with zero $2\times2$ matrix potential $V_+(x)$, \begin{equation}H_+=-I_2\partial^2,\qquad V_+(x)=\begin{pmatrix}0&0\\0&0\end{pmatrix}.\la{hamzero}\end{equation} 
Since a vector-eigenfunctions for the continuous spectrum of the new Hamiltonians $H_-$ can be straightforwardly calculated in trivial way,
\[\Psi_{\uparrow}(x;\kappa)=Q_1^-\begin{pmatrix}e^{i\kappa x}\\0\end{pmatrix},\qquad\Psi_{\downarrow}(x;\kappa)=Q_1^-\begin{pmatrix}0\\e^{i\kappa x}\end{pmatrix},\]
\[H_-\Psi_{\uparrow,\downarrow}=\kappa^2\Psi_{\uparrow,\downarrow},\qquad \kappa\in{\Bbb R},\]
so we shall seek only normalizable vector-eigenfunctions and associated vector-functions of these Hamiltonians.

It is not hard to see that for the Hamiltonian (\ref{hamzero}) there is the following complete set of linearly independent formal 
eigen- and associated of the first order vector-functions for the spectral value $\lambda=-k^2\ne0$:
\[\Psi_{1,0}(x)=\begin{pmatrix}e^{kx}\\0\end{pmatrix},\quad
\Psi_{2,0}(x)=\begin{pmatrix}e^{-kx}\\0\end{pmatrix},\quad
\Psi_{3,0}(x)=\begin{pmatrix}0\\e^{kx}\end{pmatrix},\quad
\Psi_{4,0}(x)=\begin{pmatrix}0\\e^{-kx}\end{pmatrix},\]
\[\Psi_{1,1}(x)\!=\!\begin{pmatrix}-{xe^{kx}\over{2k}}\\0\end{pmatrix},\quad
\Psi_{2,1}(x)\!=\!\begin{pmatrix}{xe^{-kx}\over{2k}}\\0\end{pmatrix},\quad
\Psi_{3,1}(x)\!=\!\begin{pmatrix}0\\-{xe^{kx}\over{2k}}\end{pmatrix},\quad
\Psi_{4,1}(x)\!=\!\begin{pmatrix}0\\{xe^{-kx}\over{2k}}\end{pmatrix},\]

\begin{equation}H_+\Psi_{i,0}=\lambda\Psi_{i,0},\qquad (H_+-\lambda I_2)\Psi_{i,1}=\Psi_{i,0},\qquad i=1,2,3,4.\la{sobprisfun}\end{equation}
These vector-functions will be used below for constructing of the intertwining operators and new Hamiltonians.

We accept the following notation in this section: $\lambda_1$ and $\lambda_2$ are eigenvalues of the matrix $\bf T$ of the intertwining operator $Q_1^-$ and $g_1^-$ is the geometric multiplicity of the eigenvalue $\lambda_1$. As well, we suppose that the matrix coefficient  $X_1^-$ at $\partial$ in the operator $Q_1^-$ is equal to the identity matrix, $X_1^-=I_2$.

\subsection{Subcase $\lambda_1\ne\lambda_2$: adding up to two bound states with different energy values\la{example1}}

In this subcase general form of transformation vector-functions $\Phi_1^-(x)$ and $\Phi_2^-(x)$ is the following in view of (\ref{sobprisfun}),
\[\Phi_1^-(x)=\begin{pmatrix}C_1e^{k_1x}+C_2e^{-k_1x}\\C_3e^{k_1x}+C_4e^{-k_1x}\end{pmatrix},\qquad \Phi_2^-(x)=\begin{pmatrix}C_5e^{k_2x}+C_6e^{-k_2x}\\C_7e^{k_2x}+C_8e^{-k_2x}\end{pmatrix},\]
\begin{equation}H_+\Phi_i^-=\lambda_j\Phi_i^-,\quad\lambda_i=-k_i^2\ne0,\qquad i=1,2,\la{pflnl}\end{equation}
where $C_1$, \dots, $C_8$ are arbitrary complex, in general, constants and we assume without the loss of generality that $C_1=1$. The remaining constants $C_2$, \dots, $C_8$ are chosen so that the Wronskian $W(x)$ of the vector-functions $\Phi_1^-(x)$ and $\Phi_2^-(x)$,
\begin{align}W(x)&=[C_7-C_3C_5]e^{(k_1+k_2)x}+[C_8-C_3C_6]e^{(k_1-k_2)x}+\nonumber\\
&+[C_2C_7-C_4C_5]e^{-(k_1-k_2)x}+[C_2C_8-C_4C_6]e^{-(k_1+k_2)x},\end{align}
does not vanish on the real axis. The operators $Q_1^-$ and $Q_1^+$, the matrix $U_0(x)$ and the new Hamiltonians $H_-$ take the following form,
\begin{align}
Q_1^\pm=\mp I_2\partial-{1\over{W(x)}}\bigg[
&\begin{pmatrix}k_1C_7-k_2C_3C_5&-(k_1-k_2)C_5\\(k_1-k_2)C_3C_7&k_2C_7-k_1C_3C_5\end{pmatrix}e^{(k_1+k_2)x}\nonumber\\
+&\begin{pmatrix}k_1C_8+k_2C_3C_6&-(k_1+k_2)C_6\\(k_1+k_2)C_3C_8&-(k_2C_8+k_1C_3C_6)\end{pmatrix}e^{(k_1-k_2)x}\nonumber\\
+&\begin{pmatrix}-(k_1C_2C_7+k_2C_4C_5)&(k_1+k_2)C_2C_5\\-(k_1+k_2)C_4C_7&(k_2C_2C_7+k_1C_4C_5)\end{pmatrix}e^{-(k_1-k_2)x}\nonumber\\
+&\begin{pmatrix}-(k_1C_2C_8-k_2C_4C_6)&(k_1-k_2)C_2C_6\\-(k_1-k_2)C_4C_8&-(k_2C_2C_8-k_1C_4C_6)\end{pmatrix}e^{-(k_1+k_2)x}
\bigg],\end{align}
\begin{align}U_0(x)={1\over{W(x)}}\bigg[
&\begin{pmatrix}-(k_1^2C_7-k_2^2C_3C_5)&(k_1^2-k_2^2)C_5\\-(k_1^2-k_2^2)C_3C_7&-(k_2^2C_7-k_1^2C_3C_5)\end{pmatrix}e^{(k_1+k_2)x}\nonumber\\
+&\begin{pmatrix}-(k_1^2C_8-k_2^2C_3C_6)&(k_1^2-k_2^2)C_6\\-(k_1^2-k_2^2)C_3C_8&-(k_2^2C_8-k_1^2C_3C_6)\end{pmatrix}e^{(k_1-k_2)x}\nonumber\\
+&\begin{pmatrix}-(k_1^2C_2C_7-k_2^2C_4C_5)&(k_1^2-k_2^2)C_2C_5\\-(k_1^2-k_2^2)C_4C_7&-(k_2^2C_2C_7-k_1^2C_4C_5)\end{pmatrix}e^{-(k_1-k_2)x}\nonumber\\
+&\begin{pmatrix}-(k_1^2C_2C_8-k_2^2C_4C_6)&(k_1^2-k_2^2)C_2C_6\\-(k_1^2-k_2^2)C_4C_8&-(k_2^2C_2C_8-k_1^2C_4C_6)\end{pmatrix}e^{-(k_1+k_2)x}
\bigg],
\end{align}
\begin{align}H_-=-&I_2\partial^2-{4\over{W^2(x)}}\nonumber\\
\times\bigg[
&\begin{pmatrix}C_3[k_1\Delta_2-k_2(\delta_2-2C_3C_5C_6)]&-k_1\Delta_2+k_2(\delta_2-2C_3C_5C_6)\\
C_3[k_1\Delta_2C_3+k_2(\delta_2C_3-2C_7C_8)]&-k_1\Delta_2C_3-k_2(\delta_2C_3-2C_7C_8)\end{pmatrix}k_2e^{2k_1x}\nonumber\\
+&\begin{pmatrix}C_7[k_2\Delta_1C_5-k_1(\delta_1C_5-2C_2C_7)]&-C_5[k_2\Delta_1C_5-k_1(\delta_1C_5-2C_2C_7)]\\
C_7[k_2\Delta_1C_7+k_1(\delta_1C_7-2C_3C_4C_5)]&-C_5[k_2\Delta_1C_7+k_1(\delta_1C_7-2C_3C_4C_5)]\end{pmatrix}k_1e^{2k_2x}\nonumber\\
+&\begin{pmatrix}-C_8[k_2\Delta_1C_6+k_1(\delta_1C_6-2C_2C_8)]&C_6[k_2\Delta_1C_6+k_1(\delta_1C_6-2C_2C_8)]\\
-C_8[k_2\Delta_1C_8-k_1(\delta_1C_8-2C_3C_4C_6)]&C_6[k_2\Delta_1C_8-k_1(\delta_1C_8-2C_3C_4C_6)]\end{pmatrix}k_1e^{-2k_2x}\nonumber\\
+&\begin{pmatrix}-C_4[k_1\Delta_2C_2+k_2(\delta_2C_2-2C_4C_5C_6)]&C_2[k_1\Delta_2C_2+k_2(\delta_2C_2-2C_4C_5C_6)]\\
-C_4[k_1\Delta_2C_4-k_2(\delta_2C_4-2C_2C_7C_8)]&C_2[k_1\Delta_2C_4-k_2(\delta_2C_4-2C_2C_7C_8)]\end{pmatrix}k_2e^{-2k_1x}\nonumber\\
+&2\begin{pmatrix}2(k_1^2C_2C_7C_8+k_2^2C_3C_4C_5C_6)&(k_1^2-k_2^2)(\delta_1C_5C_6-\delta_2C_2)\\
(k_1^2-k_2^2)(\delta_1C_7C_8-\delta_2C_3C_4)&2(k_2^2C_2C_7C_8+k_1^2C_3C_4C_5C_6)\end{pmatrix}\nonumber\\
-&\,\,\big[(k_1^2+k_2^2)\delta_1\delta_2-2k_1k_2\Delta_1\Delta_2\big]I_2
\bigg],\nonumber
\end{align}
\begin{align}\Delta_1&=C_4-C_2C_3,\qquad & \delta_1= &\,\, C_4+C_2C_3,\nonumber\\
\Delta_2&=C_5C_8-C_6C_7,\qquad & \delta_2= &\,\, C_5C_8+C_6C_7,\la{gam-62}\end{align}
so that
\begin{equation}H_+=Q_1^+Q_1^-+U_0(x),\qquad H_-=Q_1^-Q_1^++U_0(x),\qquad Q_1^-H_+=H_-Q_1^-.\la{intgl51}\end{equation}

For the spectral values $\lambda_1$ and $\lambda_2$ of the Hamiltonian $H_-$ one can easily construct formal vector-eigenfunctions
\begin{align}
\Psi_1^+(x)&=Q_1^-\begin{pmatrix}e^{k_1x}\\0\end{pmatrix}\!=\!{1\over{W(x)}}
\bigg[-(k_1\!-\!k_2)C_3\begin{pmatrix}C_5\\C_7\end{pmatrix}e^{(2k_1\!+\!k_2)x}
\!-\!(k_1\!+\!k_2)C_3\begin{pmatrix}C_6\\C_8\end{pmatrix}e^{(2k_1\!-\!k_2)x}\nonumber\\
&\qquad\quad+\begin{pmatrix}2k_1C_2C_7-(k_1-k_2)C_4C_5\\(k_1+k_2)C_4C_7\end{pmatrix}e^{k_2x}+\begin{pmatrix}2k_1C_2C_8-(k_1+k_2)C_4C_6\\(k_1-k_2)C_4C_8\end{pmatrix}e^{-k_2x}\bigg],\nonumber\\
\Psi_2^+(x)&=Q_1^-\begin{pmatrix}e^{-k_1x}\\0\end{pmatrix}\!=\!{1\over{W(x)}}
\bigg[(k_1\!-\!k_2)C_4\begin{pmatrix}C_6\\C_8\end{pmatrix}e^{-(2k_1\!+\!k_2)x}
\!+\!(k_1\!+\!k_2)C_4\begin{pmatrix}C_5\\C_7\end{pmatrix}e^{-(2k_1\!-\!k_2)x}\nonumber\\
&\qquad\quad-\begin{pmatrix}2k_1C_7-(k_1+k_2)C_3C_5\\(k_1-k_2)C_3C_7\end{pmatrix}e^{k_2x}-\begin{pmatrix}2k_1C_8-(k_1-k_2)C_3C_6\\(k_1+k_2)C_3C_8\end{pmatrix}e^{-k_2x}\bigg],\nonumber\\
\Psi_3^+(x)&=Q_1^-\begin{pmatrix}0\\e^{k_1x}\end{pmatrix}={1\over{W(x)}}
\bigg[(k_1-k_2)\begin{pmatrix}C_5\\C_7\end{pmatrix}e^{(2k_1+k_2)x}
+(k_1+k_2)\begin{pmatrix}C_6\\C_8\end{pmatrix}e^{(2k_1-k_2)x}\nonumber\\
&\qquad\quad-\begin{pmatrix}(k_1+k_2)C_2C_5\\2k_1C_4C_5-(k_1-k_2)C_2C_7\end{pmatrix}e^{k_2x}-\begin{pmatrix}(k_1-k_2)C_2C_6\\2k_1C_4C_6-(k_1+k_2)C_2C_8\end{pmatrix}e^{-k_2x}\bigg],\nonumber\\
\Psi_4^+(x)&=Q_1^-\begin{pmatrix}0\\e^{-k_1x}\end{pmatrix}\!=\!{1\over{W(x)}}
\bigg[\!-\!(k_1\!-\!k_2)C_2\!\begin{pmatrix}C_6\\C_8\end{pmatrix}\!e^{-(2k_1\!+\!k_2)x}
\!-\!(k_1\!+\!k_2)C_2\!\begin{pmatrix}C_5\\C_7\end{pmatrix}\!e^{-(2k_1\!-\!k_2)x}\nonumber\\
&\qquad\quad+\begin{pmatrix}(k_1-k_2)C_5\\2k_1C_3C_5-(k_1+k_2)C_7\end{pmatrix}e^{k_2x}+\begin{pmatrix}(k_1+k_2)C_6\\2k_1C_3C_6-(k_1-k_2)C_8\end{pmatrix}e^{-k_2x}\bigg],\nonumber\\
\Psi_5^+(x)&=Q_1^-\begin{pmatrix}e^{k_2x}\\0\end{pmatrix}\!=\!{1\over{W(x)}}
\bigg[-(k_1\!-\!k_2)C_7\begin{pmatrix}1\\C_3\end{pmatrix}e^{(k_1\!+\!2k_2)x}
\!+\!(k_1\!+\!k_2)C_7\begin{pmatrix}C_2\\C_4\end{pmatrix}e^{-(k_1\!-\!2k_2)x}\nonumber\\
&\qquad\quad-\begin{pmatrix}2k_2C_3C_6+(k_1-k_2)C_8\\(k_1+k_2)C_3C_8\end{pmatrix}e^{k_1x}-\begin{pmatrix}2k_2C_4C_6-(k_1+k_2)C_2C_8\\-(k_1-k_2)C_4C_8\end{pmatrix}e^{-k_1x}\bigg],\nonumber
\end{align}
\begin{align}
\Psi_6^+(x)&=Q_1^-\begin{pmatrix}e^{-k_2x}\\0\end{pmatrix}\!=\!{1\over{W(x)}}
\bigg[(k_1\!-\!k_2)C_8\begin{pmatrix}C_2\\C_4\end{pmatrix}e^{-(k_1\!+\!2k_2)x}
\!-\!(k_1\!+\!k_2)C_8\begin{pmatrix}1\\C_3\end{pmatrix}e^{(k_1\!-\!2k_2)x}\nonumber\\
&\qquad\quad+\begin{pmatrix}2k_2C_3C_5-(k_1+k_2)C_7\\-(k_1-k_2)C_3C_7\end{pmatrix}e^{k_1x}+\begin{pmatrix}2k_2C_4C_5+(k_1-k_2)C_2C_7\\(k_1+k_2)C_4C_7\end{pmatrix}e^{-k_1x}\bigg],\nonumber\\
\Psi_7^+(x)&=Q_1^-\begin{pmatrix}0\\e^{k_2x}\end{pmatrix}\!=\!{1\over{W(x)}}
\bigg[(k_1\!-\!k_2)C_5\begin{pmatrix}1\\C_3\end{pmatrix}e^{(k_1\!+\!2k_2)x}
\!-\!(k_1\!+\!k_2)C_5\begin{pmatrix}C_2\\C_4\end{pmatrix}e^{-(k_1\!-\!2k_2)x}\nonumber\\
&\qquad\quad+\begin{pmatrix}(k_1+k_2)C_6\\2k_2C_8+(k_1-k_2)C_3C_6\end{pmatrix}e^{k_1x}+\begin{pmatrix}-(k_1-k_2)C_2C_6\\2k_2C_2C_8-(k_1+k_2)C_4C_6\end{pmatrix}e^{-k_1x}\bigg],\nonumber\\
\Psi_8^+(x)&=Q_1^-\begin{pmatrix}0\\e^{-k_2x}\end{pmatrix}\!=\!{1\over{W(x)}}
\bigg[-(k_1\!-\!k_2)C_6\begin{pmatrix}C_2\\C_4\end{pmatrix}e^{-(k_1\!+\!2k_2)x}
\!+\!(k_1\!+\!k_2)C_6\begin{pmatrix}1\\C_3\end{pmatrix}e^{(k_1\!-\!2k_2)x}\nonumber\\
&\qquad\quad-\begin{pmatrix}-(k_1-k_2)C_5\\2k_2C_7-(k_1+k_2)C_3C_5\end{pmatrix}e^{k_1x}-\begin{pmatrix}(k_1+k_2)C_2C_5\\2k_2C_2C_7+(k_1-k_2)C_4C_5\end{pmatrix}e^{-k_1x}\bigg],\nonumber
\end{align}
\begin{equation}H_-\Psi_i^+=\lambda_1\Psi_i^+,\quad i=1,2,3,4,\qquad
H_-\Psi_j^+=\lambda_2\Psi_j^+,\quad j=5,6,7,8,\la{8sobvech}\end{equation}
only six of which are linearly independent in view of the fact that the vector-functions $\Phi_1^-(x)$ and $\Phi_2^-(x)$ (see (\ref{pflnl})) form a canonical basis in the kernel of $Q_1^-$. The latter leads to the relations
\begin{align}\Psi_1^+(x)+C_2\Psi_2^+(x)+C_3\Psi_3^+(x)+C_4\Psi_4^+(x)&=0,\nonumber\\ C_5\Psi_5^+(x)+C_6\Psi_6^+(x)+C_7\Psi_7^+(x)+C_8\Psi_8^+(x)&=0.\end{align}
It follows from the results of \cite{sokolov13} that in the considered subcase $\lambda_1\ne\lambda_2$ there is linear differential operator of the 3-rd order $Q_3^+$ with the coefficient $I_2$ at $\partial^3$ that intertwines the Hamiltonians $H_+$ and $H_-$ in the opposite direction, $Q_3^+H_-=H_+Q_3^+$, and six linearly independent vector-functions from the set (\ref{8sobvech}) form a canonical basis in the kernel of $Q_3^+$ providing an opportunity to construct $Q_3^+$ explicitly with the help of (\ref{qNrep5}). 

A linearly independent of (\ref{8sobvech}) formal vector-eigenfunctions $\Psi_9^+(x)$ and $\Psi_{10}^+(x)$ of the Hamiltonian $H_-$ for the spectral values $\lambda_1$ and $\lambda_2$ respectively can be found in the form
\begin{align}\Psi_{9}^+(x)&=Q_1^-\begin{pmatrix}-{x\over{2k_1}}e^{k_1x}+C_2{x\over{2k_1}}e^{-k_1x}\\-C_3{x\over{2k_1}}e^{k_1x}+C_4{x\over{2k_1}}e^{-k_1x}\end{pmatrix}\nonumber\\
&=-{1\over{2k_1W(x)}}\bigg[(C_7-C_3C_5)\begin{pmatrix}1\\C_3\end{pmatrix}e^{(2k_1+k_2)x}+(C_8-C_3C_6)\begin{pmatrix}1\\C_3\end{pmatrix}e^{(2k_1-k_2)x}\nonumber\\
&+2\begin{pmatrix}k_2\Delta_1C_5-k_1(\delta_1C_5-2C_2C_7)\\
k_2\Delta_1C_7\!+\!k_1(\delta_1C_7\!-\!2C_3C_4C_5)\end{pmatrix}xe^{k_2x}
-\Delta_1\begin{pmatrix}C_5\\C_7\end{pmatrix}e^{k_2x}\nonumber\\
&-\Delta_1\begin{pmatrix}C_6\\C_8\end{pmatrix}e^{-k_2x}-
2\begin{pmatrix}k_2\Delta_1C_6+k_1(\delta_1C_6-2C_2C_8)\\
k_2\Delta_1C_8\!-\!k_1(\delta_1C_8\!-\!2C_3C_4C_6)\end{pmatrix}xe^{-k_2x}\nonumber\\
&-(C_2C_7-C_4C_5)\begin{pmatrix}C_2\\C_4\end{pmatrix}e^{-(2k_1-k_2)x}-(C_2C_8-C_4C_6)\begin{pmatrix}C_2\\C_4\end{pmatrix}e^{-(2k_1+k_2)x}\bigg],\nonumber\\
\Psi_{10}^+(x)&=Q_1^-\begin{pmatrix}-C_5{x\over{2k_2}}e^{k_2x}+C_6{x\over{2k_2}}e^{-k_2x}\\-C_7{x\over{2k_2}}e^{k_2x}+C_8{x\over{2k_2}}e^{-k_2x}\end{pmatrix}\nonumber\\
&=-{1\over{2k_2W(x)}}\bigg[(C_7-C_3C_5)\begin{pmatrix}C_5\\C_7\end{pmatrix}e^{(k_1+2k_2)x}+(C_2C_7-C_4C_5)\begin{pmatrix}C_5\\C_7\end{pmatrix}e^{-(k_1-2k_2)x}\nonumber\\
&-2\begin{pmatrix}k_1\Delta_2-k_2(\delta_2-2C_3C_5C_6)\\
k_1\Delta_2C_3\!+\!k_2(\delta_2C_3\!-\!2C_7C_8)\end{pmatrix}xe^{k_1x}
+\Delta_2\begin{pmatrix}1\\C_3\end{pmatrix}e^{k_1x}\nonumber
\end{align}
\begin{align}
&+\Delta_2\begin{pmatrix}C_2\\C_4\end{pmatrix}e^{-k_1x}+
2\begin{pmatrix}k_1\Delta_2C_2\!+\!k_2(\delta_2C_2\!-\!2C_4C_5C_6)\\
k_1\Delta_2C_4\!-\!k_2(\delta_2C_4\!-\!2C_2C_7C_8)\end{pmatrix}xe^{-k_1x}\nonumber\\
&-(C_8-C_3C_6)\begin{pmatrix}C_6\\C_8\end{pmatrix}e^{(k_1-2k_2)x}-(C_2C_8-C_4C_6)\begin{pmatrix}C_6\\C_8\end{pmatrix}e^{-(k_1+2k_2)x}\bigg],\nonumber\end{align}
\begin{equation}H_-\Psi_9^+=\lambda_1\Psi_9^+,\qquad H_-\Psi^+_{10}=\lambda_2\Psi^+_{10},\la{8sobvek66}\end{equation}
since
\[(H_+-\lambda_1I_2)\begin{pmatrix}-{x\over{2k_1}}e^{k_1x}+C_2{x\over{2k_1}}e^{-k_1x}\\-C_3{x\over{2k_1}}e^{k_1x}+C_4{x\over{2k_1}}e^{-k_1x}\end{pmatrix}=\Phi_1^+(x),\]
\[(H_+-\lambda_2I_2)\begin{pmatrix}-C_5{x\over{2k_2}}e^{k_2x}+C_6{x\over{2k_2}}e^{-k_2x}\\-C_7{x\over{2k_2}}e^{k_2x}+C_8{x\over{2k_2}}e^{-k_2x}\end{pmatrix}=\Phi_2^+(x),\]
the vector-functions $\Phi_1^+(x)$ and $\Phi_2^+(x)$ belong to the kernel of $Q_1^-$ and a chain of associated vector-functions of the Hamiltonian $H_+$ is mapped (see Subsection \ref{sjio}) by the intertwining operator $Q_1^-$ into a chain of associated vector-functions of the Hamiltonian $H_-$ for the same spectral value (some first terms of the chain can be mapped by $Q_1^-$ into zeroes). 

Analysis of the vector-functions (\ref{8sobvech}) and (\ref{8sobvek66}) leads to the following results:\\

\noindent (1) if
\[{\rm{Re}}\,k_1{\rm{Re}}\,k_2>0,\qquad(C_7-C_3C_5)(C_2C_8-C_4C_6)\ne0\]
then for each of the eigenvalues $\lambda_1$ and $\lambda_2$ there is the only (up to a constant factor) normalizable vector-eigenfunction of the Hamiltonian $H_-$:
\begin{align}\Psi_{11}^+(x)&=\Psi_1^+(x)+C_3\Psi_3^+(x)=-C_2\Psi_2^+(x)-C_4\Psi_4^+(x)={1\over{W(x)}}
\nonumber\\
&\times\bigg[\!\!\begin{pmatrix}k_2\Delta_1C_5-k_1(\delta_1C_5-2C_2C_7)\\
k_2\Delta_1C_7\!+\!k_1(\delta_1C_7\!-\!2C_3C_4C_5)\end{pmatrix}e^{k_2x}\!-\!
\begin{pmatrix}k_2\Delta_1C_6+k_1(\delta_1C_6-2C_2C_8)\\
k_2\Delta_1C_8\!-\!k_1(\delta_1C_8\!-\!2C_3C_4C_6)\end{pmatrix}e^{-k_2x}\bigg],\nonumber\\
\Psi_{12}^+(x)&=C_5\Psi_5^+(x)+C_7\Psi_7^+(x)=-C_6\Psi_6^+(x)-C_8\Psi_8^+(x)={1\over{W(x)}}
\nonumber\\
&\times\bigg[\!-\!\begin{pmatrix}k_1\Delta_2-k_2(\delta_2-2C_3C_5C_6)\\
k_1\Delta_2C_3\!+\!k_2(\delta_2C_3\!-\!2C_7C_8)\end{pmatrix}e^{k_1x}\!+\!
\begin{pmatrix}k_1\Delta_2C_2\!+\!k_2(\delta_2C_2\!-\!2C_4C_5C_6)\\
k_1\Delta_2C_4\!-\!k_2(\delta_2C_4\!-\!2C_2C_7C_8)\end{pmatrix}e^{-k_1x}\bigg],\nonumber
\end{align}
\begin{equation}H_-\Psi_{11}^+=\lambda_1\Psi_{11}^+,\qquad H_-\Psi_{12}^+=\lambda_2\Psi_{12}^+,\qquad \Psi_{11}^+(x),\Psi_{12}^+(x)\in\ker Q_3^+;\la{psi910}\end{equation}
\noindent (2) if
\[{\rm{Re}}\,k_1>{\rm{Re}}\,k_2>0,\quad C_7-C_3C_5=C_5(C_4-C_2C_3)=0,\quad(C_8-C_3C_6)(C_2C_8-C_4C_6)\ne0,\]
or
\[{\rm{Re}}\,k_1>2\,{\rm{Re}}\,k_2>0,\qquad C_7-C_3C_5=0,\qquad(C_8-C_3C_6)(C_2C_8-C_4C_6)\ne0\]
or
\[{\rm{Re}}\,k_1>{\rm{Re}}\,k_2>0,\quad C_2C_8-C_4C_6=C_4-C_2C_3=0,\quad(C_7-C_3C_5)(C_2C_7-C_4C_5)\ne0,\]
or
\[{\rm{Re}}\,k_1>2\,{\rm{Re}}\,k_2>0,\qquad C_2C_8-C_4C_6=0,\qquad(C_7-C_3C_5)(C_2C_7-C_4C_5)\ne0\]
or
\[{\rm{Re}}\,k_1>2\,{\rm{Re}}\,k_2>0,\qquad\!\! C_7-C_3C_5=C_2C_8-C_4C_6=0,\qquad\!\!(C_8-C_3C_6)(C_2C_7-C_4C_5)\ne0\]
then for the eigenvalue $\lambda_1$ there is the only (up to a constant factor) normalizable vector-eigenfunction $\Psi_{11}^-(x)$ of the Hamiltonian $H_-$ and for the spectral value $\lambda_2$ there is no a normalizable vector-eigenfunction of $H_-$;\\

\noindent (3) if 
\[{\rm{Re}}\,k_1>{\rm{Re}}\,k_2>0,\quad C_7-C_3C_5=C_8-C_3C_6=0,\quad (C_2C_7-C_4C_5)(C_2C_8-C_4C_6)\ne0\]
or 
\[{\rm{Re}}\,k_1>{\rm{Re}}\,k_2>0,\quad C_2C_7-C_4C_5=C_2C_8-C_4C_6=0,\quad (C_7-C_3C_5)(C_8-C_3C_6)\ne0\]
then for the eigenvalue $\lambda_2$ there is the only (up to a constant factor) normalizable vector-eigenfunction $\Psi_{12}^-(x)$ of the Hamiltonian $H_-$ and for the spectral value $\lambda_1$ there is no a normalizable vector-eigenfunction of $H_-$;\\

\noindent (4) if 
\[2\,{\rm{Re}}\,k_2\geqslant{\rm{Re}}\,k_1>{\rm{Re}}\,k_2>0,\]
\[C_7-C_3C_5=0,\qquad C_5(C_4-C_2C_3)(C_8-C_3C_6)(C_2C_8-C_4C_6)\ne0\]
or
\[2\,{\rm{Re}}\,k_2\geqslant{\rm{Re}}\,k_1>{\rm{Re}}\,k_2>0,\]
\[C_2C_8-C_4C_6=0,\qquad (C_4-C_2C_3)(C_7-C_3C_5)(C_2C_7-C_4C_5)\ne0\]
or
\[2\,{\rm{Re}}\,k_2\geqslant{\rm{Re}}\,k_1>{\rm{Re}}\,k_2>0,\]
\[C_7-C_3C_5=C_2C_8-C_4C_6=0,\qquad (C_8-C_3C_6)(C_2C_7-C_4C_5)\ne0\]
or
\[C_7-C_3C_5=C_8-C_3C_6=C_2C_7-C_4C_5=0,\qquad C_2C_8-C_4C_6\ne0\]
or
\[C_7-C_3C_5=C_8-C_3C_6=C_2C_8-C_4C_6=0,\qquad C_2C_7-C_4C_5\ne0\]
or
\[C_7-C_3C_5=C_2C_7-C_4C_5=C_2C_8-C_4C_6=0,\qquad C_8-C_3C_6\ne0\]
or
\[C_8-C_3C_6=C_2C_7-C_4C_5=C_2C_8-C_4C_6=0,\qquad C_7-C_3C_5\ne0\]
then there is no a normalizable vector-eigenfunction of the Hamiltonian $H_-$ for the spectral values $\lambda_1$ and $\lambda_2$.\\

Let us now present some partial situations where the received formulae become significantly more simple.\\

\def\th{\,{\rm{th}}\,}

\noindent (1) For
\[C_2=1,\qquad C_3=C_4=C_5=C_6=0,\qquad C_7={1\over4}e^{-k_2x_0},\qquad C_8={1\over4}e^{k_2x_0},\]
\[{\rm{Re}}\,k_1\,{\rm{Re k_2}}\ne0,\qquad x_0\in{\Bbb R}\]
the Wronskian
\[W(x)=\ch k_1x\,\, \ch k_2(x-x_0)\]
does not have real zeroes, the operators $Q_1^+$ and $Q_1^-$, the matrix $U_0(x)$ and the new Hamiltonian $H_-$ take the form,
\begin{align}
Q_1^\pm&=\mp I_2\partial-\begin{pmatrix}k_1\th k_1x&0\\0&k_2\th k_2(x-x_0)\end{pmatrix},\nonumber\\
U_0(x)&=\begin{pmatrix}-k_1^2&0\\0&-k_2^2\end{pmatrix},\nonumber\\
H_-&=-I_2\partial^2-2\begin{pmatrix}{k_1^2\over{\ch^2k_1x}}&0\\0&{k_2^2\over{\ch^2k_2(x-x_0)}}\end{pmatrix},\nonumber\end{align}
and there are only two linearly independent vector-eigenfunctions for $H_-$,
\[\Psi_{11}^+(x)=\Psi_1^+(x)=\begin{pmatrix}{k_1\over{\ch k_1x}}\\0\end{pmatrix},\qquad\Psi_{12}^+(x)=C_7\Psi_7^+(x)={1\over4}\begin{pmatrix}0\\{k_2\over{\ch k_2(x-x_0)}}\end{pmatrix}.\]

\noindent (2) If
\[C_4=C_2C_3-{1\over{2C_6}},\qquad C_5=-C_2C_6, \qquad C_7={1\over2}-C_2C_3C_6,\qquad C_8=C_3C_6,\]
\[{\rm{Re}}\,(k_1+k_2)\ne0,\qquad C_2, C_3,C_6\in{\Bbb C},\quad C_6\ne0,\]
then the Wronskian
\[W(x)=\ch(k_1+k_2)x\]
is nonvanishing on the real axis,
\begin{align}
Q_1^\pm&=\mp\partial-{{k_1+k_2}\over{W(x)}}\bigg[C_6\begin{pmatrix}C_3&-1\\C_3^2&-C_3\end{pmatrix}e^{(k_1-k_2)x}-{1\over C_6}\begin{pmatrix}C_2C_6C_7&C_2^2C_6^2\\-C_7^2&-C_2C_6C_7\end{pmatrix}e^{-(k_1-k_2)x}\nonumber\\
&\qquad\qquad\qquad+{1\over2}\sh(k_1+k_2)x\,\,I_2\bigg]-(k_1-k_2)\begin{pmatrix}C_7-C_2C_3C_6&2C_2C_6\\2C_3C_7&-(C_7-C_2C_3C_6)\end{pmatrix}\nonumber\\
&={\bf C}\,\Bigg\{\!\!\mp\partial\!-\!{{1}\over{2W(x)}}\begin{pmatrix}k_1e^{(k_1+k_2)x}\!-\!k_2e^{-(k_1+k_2)x}&-C_6(k_1+k_2)e^{(k_1-k_2)x}\\{1\over C_6}(k_1+k_2)e^{-(k_1-k_2)x}&-k_1e^{-(k_1+k_2)x}\!+\!k_2e^{(k_1+k_2)x}\end{pmatrix}\!\!\Bigg\}{\bf C}^{-1},\nonumber\\
U_0(x)&=-{{k_1^2-k_2^2}\over{W(x)}}\bigg[C_6\begin{pmatrix}C_3&-1\\C_3^2&-C_3\end{pmatrix}e^{(k_1-k_2)x}+{1\over C_6}\begin{pmatrix}C_2C_6C_7&C_2^2C_6^2\\-C_7^2&-C_2C_6C_7\end{pmatrix}e^{-(k_1-k_2)x}\nonumber\\
&\qquad\qquad\qquad+\begin{pmatrix}C_7-C_2C_3C_6&2C_2C_6\\2C_3C_7&-(C_7-C_2C_3C_6)\end{pmatrix}\sh(k_1+k_2)x\bigg]
-{{k_1^2+k_2^2}\over2}\,I_2\nonumber\\
&={\bf C}\,\Bigg\{{1\over{2W(x)}}\!\begin{pmatrix}-k_1^2e^{(k_1+k_2)x}\!\!-\!k_2^2e^{-(k_1+k_2)x}&C_6(k_1^2-k_2^2)e^{(k_1-k_2)x}\\{1\over C_6}(k_1^2-k_2^2)e^{-(k_1-k_2)x}&-k_1^2e^{-(k_1+k_2)x}\!\!-\!k_2^2e^{(k_1+k_2)x}\end{pmatrix}\!\!\Bigg\}{\bf C}^{-1},\nonumber\\
\nonumber\\
H_-&=-I_2\partial^2-{{2(k_1+k_2)}\over{W^2(x)}}\bigg[C_6\begin{pmatrix}C_3&-1\\C_3^2&-C_3\end{pmatrix}(k_1e^{-2k_2x}-k_2e^{2k_1x})\nonumber\\
&\qquad\qquad\qquad+{1\over C_6}\begin{pmatrix}C_2C_6C_7&C_2^2C_6^2\\-C_7^2&-C_2C_6C_7\end{pmatrix}(k_1e^{2k_2x}-k_2e^{-2k_1x})+{{k_1+k_2}\over2}\,I_2
\bigg]\nonumber\\
&={\bf C}\,\Bigg\{-I_2\partial^2\!-\!{{k_1\!+\!k_2}\over{W^2(x)}}\begin{pmatrix}k_1+k_2&C_6(k_1e^{-2k_2x}\!-\!k_2e^{2k_1x})\\{1\over C_6}(k_1e^{2k_2x}\!-\!k_2e^{-2k_1x})&k_1+k_2\end{pmatrix}\Bigg\}{\bf C}^{-1},\nonumber
\end{align}
\[\Phi_1^-(x)={\bf C}\begin{pmatrix}e^{k_1x}\\-{1\over C_6}e^{-k_1x}\end{pmatrix},\qquad\Phi_2^-(x)={\bf C}\begin{pmatrix}C_6e^{-k_2x}\\e^{k_2x}\end{pmatrix},\]
\[{\bf C}=\begin{pmatrix}1&-C_2C_6\\C_3&{1\over2}-C_2C_3C_6\end{pmatrix},\qquad{\bf C}^{-1}=\begin{pmatrix}1-2C_2C_3C_6&2C_2C_6\\-2C_3&2\end{pmatrix},\qquad\det{\bf C}={1\over2}\]
and for the Hamiltonian $H_-$ there are only two linearly independent normalizable vector-eigenfunctions,
\begin{align}\Psi_{11}^+(x)&={{k_1+k_2}\over{2W(x)}}\bigg[{1\over C_6}\begin{pmatrix}C_2C_6\\-C_7\end{pmatrix}e^{k_2x}+\begin{pmatrix}1\\C_3\end{pmatrix}e^{-k_2x}\bigg]={\bf C}\,\Bigg\{{{k_1+k_2}\over{2}}\begin{pmatrix}{e^{-k_2x}\over{\ch(k_1+k_2)x}}\\-{1\over C_6}{e^{k_2x}\over{\ch(k_1+k_2)x}}\end{pmatrix}\Bigg\},\nonumber\\
\Psi_{12}^+(x)&={{k_1+k_2}\over{2W(x)}}\bigg[C_6\begin{pmatrix}1\\C_3\end{pmatrix}e^{k_1x}-\begin{pmatrix}C_2C_6\\
-C_7\end{pmatrix}
e^{-k_1x}\bigg]={\bf C}\,\Bigg\{{{k_1+k_2}\over2}\begin{pmatrix}C_6{e^{k_1x}\over{\ch(k_1+k_2)x}}\\{e^{-k_1x}\over{\ch(k_1+k_2)x}}\end{pmatrix}\Bigg\},\nonumber
\end{align}

\[{\rm{Re}}\,k_1\,{\rm{Re}}\,k_2>0,\]
or the only (up to a constant factor) normalizable vector-eigenfunction $\Psi_{11}^+(x)$ for
\[{\rm{Re}}\,k_1\,{\rm{Re}}\,k_2\leqslant0,\qquad|{\rm{Re}}\,k_1|>2|{\rm{Re}}\,k_2|\] 
or the only (up to a constant factor) normalizable vector-eigenfunction $\Psi_{12}^+(x)$ for
\[{\rm{Re}}\,k_1\,{\rm{Re}}\,k_2\leqslant0,\qquad|{\rm{Re}}\,k_2|>2|{\rm{Re}}\,k_1|\] 
or there is no a normalizable vector-eigenfunction for
\[{\rm{Re}}\,k_1\,{\rm{Re}}\,k_2\leqslant0,\qquad4|{\rm{Re}}\,k_2|\geqslant2|{\rm{Re}}\,k_1|\geqslant|{\rm{Re}}\,k_2|.\]
It is evident here and in the what follows below in this Subsection \ref{example1} that the representations and the intertwining (\ref{intgl51}) transform trivially into the analogous formulae for the Hamiltonians ${\bf C}^{-1}H_+{\bf C}=H_+=-\partial^2$ and ${\bf C}^{-1}H_-{\bf C}$, for the matrix ${\bf C}^{-1}U_0(x){\bf C}$ and for the operators ${\bf C}^{-1}Q_1^+{\bf C}$ and ${\bf C}^{-1}Q_1^-{\bf C}$, that ${\bf C}^{-1}\Psi_{11}^+(x)$ and ${\bf C}^{-1}\Psi_{12}^+(x)$ are vector-eigenfunctions (formal sometimes) of the Hamiltonian ${\bf C}^{-1}H_-{\bf C}$ for the same eigenvalues $\lambda_1=-k_1^2$ and $\lambda_2=-k_2^2$ respectively and that ${\bf C}^{-1}\Phi_1^-(x)$ and ${\bf C}^{-1}\Phi_2^-(x)$ are transformation vector-functions corresponding to conversion of the Hamiltonian ${\bf C}^{-1}H_+{\bf C}=H_+$ to the Hamiltonian ${\bf C}^{-1}H_-{\bf C}$ with the help of the intertwining operator ${\bf C}^{-1}Q_1^-{\bf C}$.\\

\noindent (3) For
\[C_2=\alpha C_5,\quad C_4=\alpha\Big({1\over 2}+C_3C_5\Big),\quad C_6=C_5,\quad C_7={1\over2}+C_3C_5,\quad C_8={1\over2}+C_3C_5,\]
\[{\rm{Re}}\,k_2\ne0,\qquad \alpha,C_3, C_5\in{\Bbb C},\]
the Wronskian
\[W(x)=e^{k_1x}\ch k_2x\]
does not vanish on the real axis as well, we have
\begin{align}
Q_1^\pm&=\!\mp I_2\partial\!+\!\!2
k_2\!\begin{pmatrix}\!C_3C_5\!&\!\!-C_5\!\\\!C_3C_7\!&\!\!-C_7\!\end{pmatrix}\!\!\th k_2x\!\!
+\!\!2\alpha\!\begin{pmatrix}\!C_5C_7\!\!\!&-C_5^2\\C_7^2&\!\!\!-C_5C_7\!\end{pmatrix}\!\!\Big[\!k_1\!\!+\!\!k_2\th k_2x\!\Big]e^{\!-\!2k_1x}\!
\!\!-\!2k_1\!\!\begin{pmatrix}C_7&-C_5\\\!C_3C_7\!&\!\!\!-C_3C_5\!\end{pmatrix}\nonumber\\
&={\bf C}\,\bigg\{\mp I_2\partial+
\begin{pmatrix}-k_1&0\\ \alpha\big[k_1+k_2\th k_2x\big]e^{-2k_1x}&-k_2\th k_2x\end{pmatrix}\bigg\}\,{\bf C}^{-1},\nonumber\\
\nonumber\end{align}
\begin{align}
U_0(x)&=-(k_1^2\!-\!k_2^2)\bigg[2\alpha\!\begin{pmatrix}C_5C_7\!\!&-C_5^2\\C_7^2&\!\!-C_5C_7\end{pmatrix}e^{-2k_1x}
\!+\!\begin{pmatrix}C_7\!+\!C_3C_5\!\!&-2C_5\\2C_3C_7&\!\!-(C_7\!+\!C_3C_5)\end{pmatrix}\!\bigg]\!-\!{{k_1^2\!+\!k_2^2}\over2}\,I_2
\nonumber\\
&={\bf C}\begin{pmatrix}-k_1^2&0\\-\alpha(k_1^2-k_2^2) e^{-2k_1x}&-k_2^2\end{pmatrix}{\bf C}^{-1},\nonumber\\
\nonumber\\
H_-&=\!-I_2\partial^2\!\!+\!4\bigg\{\!\!
\begin{pmatrix}C_3C_5\!&\!\!-C_5\\
C_3C_7\!&\!\!-C_7\end{pmatrix}\!{k_2^2\over{\!\!\ch^2 k_2x\!\!}}\!+\!\alpha\!\begin{pmatrix}C_5C_7\!\!\!&-C_5^2\\
C_7^2&\!\!\!-C_5C_7\end{pmatrix}\!\!\bigg[{k_2^2\over{\!\ch^2k_2x\!\!}}\!-\!2k_1^2\!-\!2k_1k_2\th k_2x\bigg]e^{-2k_1x}\!
\bigg\}\nonumber\\
&={\bf C}\,\Bigg\{-I_2\partial^2+
\begin{pmatrix}0&0\\
\alpha\big[{{2k_2^2}\over{\ch^2k_2x}}-4k_1^2 - 4k_1k_2\th k_2x\big]e^{-2k_1x}&-{{2k_2^2}\over{\ch^2 k_2x}}\end{pmatrix}
\Bigg\}{\bf C}^{-1},\nonumber\\
\nonumber
\end{align}
\[\Phi_1^-(x)={\bf C}\begin{pmatrix}e^{k_1x}\\ \alpha e^{-k_1x}\end{pmatrix},\qquad\Phi_2^-(x)={\bf C}\begin{pmatrix}0\\2\ch k_2x\end{pmatrix},\]
\begin{align}
\Psi_{11}^+(x)&=\alpha\begin{pmatrix}C_5\\C_7\end{pmatrix}\big[k_1+k_2\th k_2x\big]e^{-k_1x}={\bf C}\begin{pmatrix}0\\ \alpha[k_1+k_2\th k_2x]e^{-k_1x}\end{pmatrix},
\nonumber\\
\Psi_{12}^+(x)&=\begin{pmatrix}C_5\\C_7\end{pmatrix}{k_2\over{\ch k_2x}}={\bf C}\begin{pmatrix}0\\{k_2\over{\ch k_2x}}\end{pmatrix},\nonumber
\end{align}
\[{\bf C}=\begin{pmatrix}1&C_5\\C_3&{1\over2}+C_3C_5\end{pmatrix},\qquad{\bf C}^{-1}=\begin{pmatrix}1+2C_3C_5&-2C_5\\-2C_3&2\end{pmatrix},\qquad\det{\bf C}={1\over2}\]
and for the new Hamiltonian $H_-$ there is the only (up to a constant factor) normalizable vector-eigenfunction $\Psi_{12}^+(x)$.\\

\noindent (4) If
\[C_2=0,\qquad C_4=0,\qquad C_7=1+C_3C_5,\qquad C_8=C_3C_6,\qquad C_3,C_5,C_6\in{\Bbb C},\]
then the Wronskian 
\[W(x)=e^{(k_1+k_2)x}\]
is without real zeroes again,
\begin{align}
Q_1^\pm&=\mp I_2\partial\!-\!(k_1\!+\!k_2)C_6\!\begin{pmatrix}C_3&-1\\C_3^2&\!\!-C_3\end{pmatrix}\!e^{-2k_2x}\!\!\!-\!{{k_1\!-\!k_2}\over2}\!\begin{pmatrix} C_7\!+\!C_3C_5\!\!&-2C_5\\2C_3C_7&\!\!-(C_7\!+\!C_3C_5)\end{pmatrix}\!-\!{{k_1\!+\!k_2}\over2}I_2
\nonumber\\
&={\bf C}\,\bigg\{\mp I_2\partial+\begin{pmatrix}-k_1&(k_1+k_2)C_6e^{-2k_2x}\\0&-k_2\end{pmatrix}\bigg\}{\bf C}^{-1},
\nonumber\\
U_0(x)&=-(k_1^2\!-\!k_2^2)C_6\begin{pmatrix}C_3&-1\\C_3^2&-C_3\end{pmatrix}e^{-2k_2x}\!-\!{{k_1^2\!-\!k_2^2}\over2}
\begin{pmatrix}C_7\!+\!C_3C_5&-2C_5\\2C_3C_7&-(C_7\!+\!C_3C_5)\end{pmatrix}\!-\!{{k_1^2\!+\!k_2^2}\over2}I_2\nonumber\\
&={\bf C}\begin{pmatrix}-k_1^2&(k_1^2-k_2^2)C_6e^{-2k_2x}\\0&-k_2^2\end{pmatrix}{\bf C}^{-1},
\nonumber\\
H_-&=-I_2\partial^2+4k_2(k_1+k_2)C_6\begin{pmatrix}C_3&-1\\
C_3^2&-C_3\end{pmatrix}e^{-2k_2x}\nonumber\\
&={\bf C}\,\bigg\{-I_2\partial^2+\begin{pmatrix}0&-4k_2(k_1+k_2)C_6e^{-2k_2x}\\
0&0\end{pmatrix}\bigg\}{\bf C}^{-1},
\nonumber
\end{align}
\[\Phi_1^-(x)={\bf C}\begin{pmatrix}e^{k_1x}\\0\end{pmatrix},\qquad
\Phi_2^-(x)={\bf C}\begin{pmatrix}C_6e^{-k_2x}\\e^{k_2x}\end{pmatrix},\]
\begin{align}\Psi_{11}^+(x)&=\begin{pmatrix}0\\0\end{pmatrix}\!=\!{\bf C}\begin{pmatrix}0\\0\end{pmatrix},\quad
\Psi_{12}^+(x)\!=\!(k_1\!+\!k_2)C_6\begin{pmatrix}1\\C_3\end{pmatrix}e^{-k_2x}\!=\!{\bf C}\begin{pmatrix}(k_1\!+\!k_2)C_6e^{-k_2x}\\0\end{pmatrix},\nonumber
\end{align}
\[{\bf C}=\begin{pmatrix}1&C_5\\C_3&1+C_3C_5\end{pmatrix},\qquad{\bf C}^{-1}=\begin{pmatrix}1+C_3C_5&-C_5\\-C_3&1\end{pmatrix},\qquad\det{\bf C}=1\]
and for the Hamiltonian $H_-$ there is no a normalizable vector-eigenfunction.\\

In general, the formulae (\ref{pflnl})--(\ref{gam-62}) and (\ref{psi910}) can be simplified with the help of similarity transformation for $\Delta_1\ne0\Leftrightarrow C_4\ne C_2C_3$ as follows,
\[{\bf C}^{-1}\Phi_1^-(x)=\begin{pmatrix}e^{k_1x}\\e^{-k_1x}\end{pmatrix},\qquad{\bf C}^{-1}\Phi_2^-(x)=\begin{pmatrix}\tilde C_5e^{k_2x}+\tilde C_6e^{-k_2x}\\\tilde C_7e^{k_2x}+\tilde C_8e^{-k_2x}\end{pmatrix},\]
\[\tilde W(x)=\tilde C_7e^{(k_1+k_2)x}+\tilde C_8 e^{(k_1-k_2)x}-\tilde C_5e^{-(k_1-k_5)x}-\tilde C_6e^{-(k_1+k_2)x}={1\over\Delta_1}W(x),\]
\begin{align}{\bf C}^{-1}Q_1^\pm{\bf C}&=\mp I_2\partial-\begin{pmatrix}k_1&0\\0&-k_1\end{pmatrix}-{1\over{\tilde W(x)}}\bigg[
(k_1+k_2)\begin{pmatrix}\tilde C_6e^{-(k_1+k_2)x}&-\tilde C_6e^{(k_1-k_2)x}\\-\tilde C_7e^{-(k_1-k_2)x}&\tilde C_7e^{(k_1+k_2)x}\end{pmatrix}\nonumber\\
&+(k_1-k_2)\begin{pmatrix}\tilde C_5e^{-(k_1-k_2)x}&-\tilde C_5e^{(k_1+k_2)x}\\-\tilde C_8e^{-(k_1+k_2)x}&\tilde C_8e^{(k_1-k_2)x}\end{pmatrix}\bigg],\nonumber\end{align}
\begin{align}{\bf C}^{-1}U_0(x){\bf C}=-k_1^2I_2\!-\!{{\!(k_1^2\!-\!k_2^2)\!}\over{\tilde W(x)}}\!
\begin{pmatrix}\tilde C_5e^{\!-(k_1\!-\!k_2)x}\!\!\!+\!\tilde C_6e^{\!-(k_1\!+\!k_2)x}\!&\!-\tilde C_5e^{(k_1\!+\!k_2)x}\!\!\!-\!\tilde C_6e^{(k_1\!-\!k_2)x}\\
\tilde C_7e^{\!-(k_1\!-\!k_2)x}\!\!\!+\!\tilde C_8e^{\!-(k_1\!+\!k_2)x}\!&\!\!-\tilde C_7e^{(k_1\!+\!k_2)x}\!\!\!-\!\tilde C_8e^{(k_1\!-\!k_2)x}\end{pmatrix},\nonumber
\end{align}
\begin{align}{\bf C}^{-1}H_-{\bf C}&=-I_2\partial^2\!-\!{{4}\over{\tilde W^2(x)}}
\bigg[2k_2^2\begin{pmatrix}\tilde C_5\tilde C_6e^{-2k_1x}\!\!&-\tilde C_5\tilde C_6\\-\tilde C_7\tilde C_8&\!\!\tilde C_7\tilde C_8e^{2k_1x}\end{pmatrix}\nonumber\\
&+k_2[(k_1-k_2)\tilde C_5\tilde C_8-(k_1+k_2)\tilde C_6\tilde C_7]\begin{pmatrix}1&-e^{2k_1x}\\-e^{-2k_1x}&1\end{pmatrix}
\nonumber\\
&+k_1k_2\begin{pmatrix}[\tilde C_5e^{k_2x}\!\!-\!\tilde C_6e^{\!-k_2x}][\tilde C_7e^{k_2x}\!\!+\!\tilde C_8e^{\!-k_2x}]&
-[\tilde C_5e^{k_2x}\!\!+\!\tilde C_6e^{\!-k_2x}][\tilde C_5e^{k_2x}\!\!-\!\tilde C_6e^{\!-k_2x}]\\
[\tilde C_7e^{k_2x}\!\!-\!\tilde C_8e^{\!-k_2x}][\tilde C_7e^{k_2x}\!\!+\!\tilde C_8e^{\!-k_2x}]&
-[\tilde C_5e^{k_2x}\!\!+\!\tilde C_6e^{\!-k_2x}][\tilde C_7e^{k_2x}\!\!-\!\tilde C_8e^{\!-k_2x}]\end{pmatrix}\nonumber\\
&-k_1^2\begin{pmatrix}[\tilde C_5e^{k_2x}\!\!+\!\tilde C_6e^{\!-k_2x}][\tilde C_7e^{k_2x}\!\!+\!\tilde C_8e^{\!-k_2x}]\!\!&-[\tilde C_5e^{k_2x}+\tilde C_6e^{-k_2x}]^2\\
-[\tilde C_7e^{k_2x}+\tilde C_8e^{-k_2x}]^2&\!\![\tilde C_5e^{k_2x}\!\!+\!\tilde C_6e^{\!-k_2x}][\tilde C_7e^{k_2x}\!\!+\!\tilde C_8e^{\!-k_2x}]\end{pmatrix}\bigg],
\nonumber
\end{align}
\begin{align}{\bf C}^{-1}\Psi_{11}^+(x)&={1\over{\tilde W(x)}}\!\bigg[-\begin{pmatrix}(k_1-k_2)\tilde C_5\\
-(k_1+k_2)\tilde C_7\end{pmatrix}e^{k_2x}-
\begin{pmatrix}(k_1+k_2)\tilde C_6\\-(k_1-k_2)\tilde C_8\end{pmatrix}e^{-k_2x}\bigg],\nonumber\\
{\bf C}^{-1}\Psi_{12}^+(x)&={1\over{\tilde W(x)}}\!\bigg[\!\!\begin{pmatrix}\!(k_1\!+\!k_2)\tilde C_6\tilde C_7\!-\!(k_1\!-\!k_2)
\tilde C_5\tilde C_8\!\\2k_2\tilde C_7\tilde C_8\end{pmatrix}\!\!e^{k_1x}\!\!\!-\!
\begin{pmatrix}2k_2\tilde C_5\tilde C_6\\\!(k_1\!+\!k_2)\tilde C_6\tilde C_7\!-\!(k_1\!-\!k_2)\tilde C_5\tilde C_8\!\end{pmatrix}\!\!e^{-k_1x}\!\bigg],\nonumber
\end{align}
\[{\bf C}=\begin{pmatrix}1&C_2\\C_3&C_4\end{pmatrix},\qquad{\bf C}^{-1}={1\over{\Delta_1}}\begin{pmatrix}C_4&-C_2\\-C_3&1\end{pmatrix},\qquad \det{\bf C}=\Delta_1,\]
\[\tilde C_5=-{{C_2C_7-C_4C_5}\over\Delta_1},\quad\tilde C_6=-{{C_2C_8-C_4C_6}\over\Delta_1},\quad\tilde C_7={{C_7-C_3C_5}\over\Delta_1},\quad\tilde C_8={{C_8-C_3C_6}\over\Delta_1}\]
and for $\Delta_1=0\Leftrightarrow C_4= C_2C_3$ as follows,
\[{\bf C}^{-1}\Phi_1^-(x)=\begin{pmatrix}e^{k_1x}+C_2e^{-k_1x}\\0\end{pmatrix},\qquad{\bf C}^{-1}\Phi_2^-(x)=\begin{pmatrix}C_5e^{k_2x}+C_6e^{-k_2x}\\\tilde C_7e^{k_2x}+\tilde C_8e^{-k_2x}\end{pmatrix},\]
\begin{align}\tilde W(x)&=\big[e^{k_1x}+C_2e^{-k_1x}\big]\big[\tilde C_7e^{k_2x}+\tilde C_8e^{-k_2x}\big]=-{1\over\alpha}W(x),\nonumber\\
{\bf C}^{-1}Q_1^\pm{\bf C}&=\mp I_2\partial
\!-\!\begin{pmatrix}k_1{{e^{k_1x}\!-\!C_2e^{-k_1x}}\over{e^{k_1x}\!+\!C_2e^{-k_1x}}}&
k_2{{C_5e^{k_2x}\!-\!C_6e^{-k_2x}}\over{\tilde C_7e^{k_2x}\!+\!\tilde C_8e^{-k_2x}}}
\!-\!k_1{{[e^{k_1x}\!-\!C_2e^{-k_1x}][C_5e^{k_2x}\!+\!C_6e^{-k_2x}]}\over
{[e^{k_1x}\!+\!C_2e^{-k_1x}][\tilde C_7e^{k_2x}\!+\!\tilde C_8e^{-k_2x}]}}
\\0&k_2{{\tilde C_7e^{k_2x}-\tilde C_8e^{-k_2x}}\over{\tilde C_7e^{k_2x}+\tilde C_8e^{-k_2x}}}\end{pmatrix},\nonumber\\
{\bf C}^{-1}U_0(x){\bf C}&=-\begin{pmatrix}k_1^2&-(k_1^2-k_2^2){{C_5e^{k_2x}+C_6e^{-k_2x}}\over{\tilde C_7e^{k_2x}+\tilde C_8e^{-k_2x}}}\\0&k_2^2\end{pmatrix},\nonumber\\
{\bf C}^{-1}H_-{\bf C}&=-I_2\partial^2-\begin{pmatrix}{{8k_1^2C_2}\over{[e^{k_1x}+C_2e^{-k_1x}]^2}}&-{{8k_1^2C_2[C_5e^{k_2x}+C_6e^{-k_2x}]}\over{[e^{k_1x}+C_2e^{-k_1x}]^2[\tilde C_7e^{k_2x}+\tilde C_8e^{-k_2x}]}}
\\0&{{8k_2^2\tilde C_7\tilde C_8}\over{[\tilde C_7e^{k_2x}+\tilde C_8e^{-k_2x}]^2}}\end{pmatrix}\nonumber\\
&+\begin{pmatrix}0&{{4k_1k_2(C_5\tilde C_8-C_6\tilde C_7)[e^{k_1x}-C_2e^{-k_1x}]}\over{[e^{k_1x}+C_2e^{-k_1x}][\tilde C_7e^{k_2x}+\tilde C_8e^{-k_2x}]^2}}-{{4k_2^2(C_5\tilde C_8+C_6\tilde C_7)}\over{[\tilde C_7e^{k_2x}+\tilde C_8e^{-k_2x}]^2}}
\\0&0\end{pmatrix},\nonumber\\
{\bf C}^{-1}\Psi_{11}^+(x)&=\begin{pmatrix}{{2k_1C_2}\over{e^{k_1x}\!+\!C_2e^{-k_1x}}}\\
0\end{pmatrix}\!,\quad
{\bf C}^{-1}\Psi_{12}^+(x)\!=\!\begin{pmatrix}{{k_2(C_5\tilde C_8\!+\!C_6\tilde C_7)}\over{\tilde C_7e^{k_2x}\!+\!\tilde C_8e^{-k_2x}}}\!-\!
{{k_1(C_5\tilde C_8-C_6\tilde C_7)[e^{k_1x}-C_2e^{-k_1x}]}\over{[e^{k_1x}\!+\!C_2e^{-k_1x}][\tilde C_7e^{k_2x}\!+\!\tilde C_8e^{-k_2x}]}}\\
{{2k_2\tilde C_7\tilde C_8}\over{\tilde C_7e^{k_2x}+\tilde C_8e^{-k_2x}}}\end{pmatrix}\!,\nonumber
\end{align}
\[{\bf C}=\begin{pmatrix}1&0\\C_3&-\alpha\end{pmatrix},\qquad{\bf C}^{-1}=\begin{pmatrix}1&0\\{C_3\over\alpha}&-{1\over\alpha}\end{pmatrix},\qquad\det{\bf C}=-\alpha,\qquad\alpha\in{\Bbb C},\,\,\,\alpha\ne0,\]
\[\tilde C_7=-{1\over\alpha}(C_7-C_3C_5),\qquad\tilde C_8=-{1\over\alpha}(C_8-C_3C_6),\]
where $\tilde W(x)$ is the Wronskian of ${\bf C}^{-1}\Phi_1^-(x)$ and ${\bf C}^{-1}\Phi_2^-(x)$.

\subsection{Subcase $\lambda_1\!=\!\lambda_2$, $g^-_1\!=\!2$: adding up to two bound states de\-s\-cri\-bed by vector-eigenfunctions with the same energy value\la{example2}}

In this subcase the formulae (\ref{pflnl}) -- (\ref{psi910}) are still valid with $k_1=k_2$ and we additionally to the condition $C_1=1$ assume without the loss of generality that $C_5=0$, since the latter condition can be achieved in any case by the change of a canonical basis in the kernel of $Q_1^-$: $\Phi_1^-(x)$ and $\Phi_2^-(x)-C_5\Phi_1^-(x)$ instead of $\Phi_1^-(x)$ and $\Phi_2^-(x)$. Thus, the formulae (\ref{pflnl}) -- (\ref{psi910}) take in the considered subcase the following more simple form:
\[\Phi_1^-(x)=\begin{pmatrix}C_1e^{kx}+C_2e^{-kx}\\C_3e^{kx}+C_4e^{-kx}\end{pmatrix},\quad \Phi_2^-(x)=\begin{pmatrix}C_5e^{kx}+C_6e^{-kx}\\C_7e^{kx}+C_8e^{-kx}\end{pmatrix},\qquad C_1=1, \quad C_5=0,\]
\begin{equation}H_+\Phi_i^-=\lambda\Phi_i^-,\quad i=1,2,\qquad\lambda=\lambda_1=\lambda_2=-k^2\ne0,\qquad k=k_1=k_2,\la{pflrl}\end{equation}
\begin{equation}W(x)=C_7e^{2kx}+[C_8-C_3C_6+C_2C_7]+[C_2C_8-C_4C_6]e^{-2kx},\la{W68}\end{equation}
\begin{align}
Q_1^\pm=\mp I_2\partial-&{k\over{W(x)}}\bigg\{\Big[
C_7e^{2kx}-(C_2C_8-C_4C_6)e^{-2kx}\Big]I_2\nonumber\\
+&\begin{pmatrix}(C_8+C_3C_6-C_2C_7)&-2C_6\\2(C_3C_8-C_4C_7)&-(C_8+C_3C_6-C_2C_7)\end{pmatrix}
\bigg\},\la{qpr2}\end{align}
\begin{equation}U_0(x)=-k^2I_2,\end{equation}
\begin{align}H_-=-I_2\partial^2-{{8k^2}\over{W^2(x)}}\bigg[&
C_7\begin{pmatrix}C_2C_7-C_3C_6&C_6\\
-(C_3C_8-C_4C_7)&C_8\end{pmatrix}e^{2kx}\nonumber\\
+&(C_2C_8-C_4C_6)\begin{pmatrix}C_8&-C_6\\
C_3C_8-C_4C_7&C_2C_7-C_3C_6\end{pmatrix}e^{-2kx}\nonumber\\
+&2C_7(C_2C_8-C_4C_6)I_2
\bigg],\la{H71}
\end{align}
\begin{align}
\Psi_1^+(x)&=\Psi_5^+(x)=Q_1^-\begin{pmatrix}e^{kx}\\0\end{pmatrix}\!=\!{{2k}\over{W(x)}}
\bigg[\begin{pmatrix}C_2C_7-C_3C_6\\-(C_3C_8\!-\!C_4C_7)\end{pmatrix}e^{kx}\!\!+\!(C_2C_8\!-\!C_4C_6)\begin{pmatrix}1\\0\end{pmatrix}e^{-kx}\bigg],\nonumber\\
\Psi_2^+(x)&=\Psi_6^+(x)=Q_1^-\begin{pmatrix}e^{-kx}\\0\end{pmatrix}\!=\!{{2k}\over{W(x)}}
\bigg[-C_7\begin{pmatrix}1\\0\end{pmatrix}e^{kx}-\begin{pmatrix}C_8\\C_3C_8-C_4C_7\end{pmatrix}e^{-kx}\bigg],\nonumber\\
\Psi_3^+(x)&=\Psi_7^+(x)=Q_1^-\begin{pmatrix}0\\e^{kx}\end{pmatrix}={{2k}\over{W(x)}}
\bigg[\begin{pmatrix}C_6\\C_8\end{pmatrix}e^{kx}+(C_2C_8-C_4C_6)\begin{pmatrix}0\\1\end{pmatrix}e^{-kx}\bigg],\nonumber\\
\Psi_4^+(x)&=\Psi_8^+(x)=Q_1^-\begin{pmatrix}0\\e^{-kx}\end{pmatrix}\!=\!{{2k}\over{W(x)}}
\bigg[-C_7\begin{pmatrix}0\\1\end{pmatrix}e^{kx}-\begin{pmatrix}-C_6\\C_2C_7-C_3C_6\end{pmatrix}e^{-kx}\bigg],
\la{4sobvech}\end{align}
\begin{align}\Psi_{9}^+(x)&=Q_1^-\begin{pmatrix}-{x\over{2k}}e^{kx}+C_2{x\over{2k}}e^{-kx}\\-C_3{x\over{2k}}e^{kx}+C_4{x\over{2k}}e^{-kx}\end{pmatrix}= -{1\over{2kW(x)}}\bigg[C_7\begin{pmatrix}1\\C_3\end{pmatrix}e^{3kx}+4kC_7\begin{pmatrix}C_2\\
C_4\end{pmatrix}xe^{kx}\nonumber\\
&+\begin{pmatrix}C_8-C_3C_6\\C_3(C_8-C_3C_6)-C_7(C_4-C_2C_3)\end{pmatrix}e^{kx}
-\begin{pmatrix}C_2^2C_7+C_6(C_4-C_2C_3)\\C_2C_4C_7+C_8(C_4-C_2C_3)\end{pmatrix}e^{-kx}\nonumber\\
&+4k(C_2C_8-C_4C_6)\begin{pmatrix}1\\
C_3\end{pmatrix}xe^{-kx}-(C_2C_8-C_4C_6)\begin{pmatrix}C_2\\C_4\end{pmatrix}e^{-3kx}\bigg],\nonumber\\ 
\Psi_{10}^+(x)&=Q_1^-\begin{pmatrix}C_6{x\over{2k}}e^{-kx}\\-C_7{x\over{2k}}e^{kx}+C_8{x\over{2k}}e^{-kx}\end{pmatrix}=-{1\over{2kW(x)}}\bigg[C_7\begin{pmatrix}0\\C_7\end{pmatrix}e^{3kx}+4kC_7
\begin{pmatrix}C_6\\C_8\end{pmatrix}xe^{kx}\nonumber\\
&-C_7\begin{pmatrix}C_6\\C_3C_6-C_2C_7\end{pmatrix}e^{kx}-\begin{pmatrix}C_6(C_8-C_3C_6)+C_2C_6C_7\\C_8(C_8-C_3C_6)+C_4C_6C_7\end{pmatrix}e^{-kx}\nonumber\\
&+4k(C_2C_8-C_4C_6)\begin{pmatrix}0\\C_7\end{pmatrix}xe^{-kx}-(C_2C_8-C_4C_6)\begin{pmatrix}C_6\\C_8\end{pmatrix}e^{-3kx}\bigg],\la{4sobvek74}\end{align}
\begin{align}
\Psi_{11}^+(x)&=\Psi_1^+(x)+C_3\Psi_3^+(x)=-C_2\Psi_2^+(x)-C_4\Psi_4^+(x)
\nonumber\\
&\qquad\qquad\qquad\qquad={{2k}\over{W(x)}}\bigg[C_7\begin{pmatrix}C_2\\
C_4\end{pmatrix}e^{kx}+(C_2C_8-C_4C_6)
\begin{pmatrix}1\\C_3\end{pmatrix}e^{-kx}\bigg],\nonumber\\
\Psi_{12}^+(x)&=C_5\Psi_5^+(x)+C_7\Psi_7^+(x)=-C_6\Psi_6^+(x)-C_8\Psi_8^+(x)
\nonumber\\
&\qquad\qquad\qquad\qquad=C_7{{2k}\over{W(x)}}\bigg[\begin{pmatrix}C_6\\
C_8\end{pmatrix}e^{kx}+(C_2C_8-C_4C_6)
\begin{pmatrix}0\\1\end{pmatrix}e^{-kx}\bigg],\nonumber
\end{align}
\begin{equation}H_-\Psi_i^+=\lambda\Psi_i^+,\quad i=1,\ldots,12,\la{psi73}\end{equation}
\begin{align}\Psi_1^+(x)+C_2\Psi_2^+(x)+C_3\Psi_3^+(x)+C_4\Psi_4^+(x)&=0,\nonumber\\ C_6\Psi_2^+(x)+C_7\Psi_3^+(x)+C_8\Psi_4^+(x)&=0,\la{links74}\end{align}
where the constants $C_2$, $C_3$, $C_4$, $C_6$, $C_7$ and $C_8$ are chosen so that the Wronskian (\ref{W68}) does not have real zeroes. Moreover, the relations (\ref{intgl51}) in accordance  with the results of Section \ref{SO1O} can be supplemented by the additional intertwining relation with the operator $Q_1^+$ as follows,
\begin{equation}H_+=Q_1^+Q_1^-+U_0(x),\,\,\, H_-=Q_1^-Q_1^++U_0(x),\quad Q_1^-H_+=H_-Q_1^-,\,\,\, Q_1^+H_-=H_+Q_1^+.\la{dopsootn75}\end{equation}

There are two only linearly independent vector-functions in the set $\Psi_1^+(x)$,  $\Psi_2^+(x)$, $\Psi_3^+(x)$ and $\Psi_4^+(x)$ in view of the fact that the vector-functions $\Phi_1^-(x)$ and $\Phi_2^-(x)$ (see (\ref{pflrl})) form a canonical basis in the kernel of $Q_1^-$. The corresponding relations between the vector-functions $\Psi_1^+(x)$,  $\Psi_2^+(x)$, $\Psi_3^+(x)$ and $\Psi_4^+(x)$ are expressed by the formulae (\ref{links74}). It is nod hard to check that two linearly independent of these vector-functions form a canonical basis in the kernel of the intertwining operator $Q_1^+$ and that these two vector-functions draw up together with the vector-functions $\Psi_9^+(x)$ and $\Psi_{10}^+(x)$ a complete set of linearly independent formal vector-eigenfunctions of the Hamiltonian $H_-$ for the spectral value $\lambda$. In addition, the vector-functions $\Psi_{11}^+(x)$ and $\Psi_{12}^+(x)$ as a linear combinations of $\Psi_1^+(x)$,  $\Psi_2^+(x)$, $\Psi_3^+(x)$ and $\Psi_4^+(x)$ belong to the kernel of $Q_1^+$.

Analysis of the vector-functions (\ref{4sobvech}), (\ref{4sobvek74}) and (\ref{psi73}) leads to the following results:\\

\noindent (1) if
\[{\rm{Re}}\,k\ne0,\qquad C_7(C_2C_8-C_4C_6)\ne0\]
then for the eigenvalue $\lambda$ there are only two linearly independent normalizable vector-eigenfunctions of the Hamiltonian $H_-$:
\[\Psi_1^+(x),\quad \Psi_3^+(x)\qquad\text{or}\qquad\Psi_2^+(x),\quad\Psi_4^+(x)\qquad\text{or}\qquad\Psi_{11}^+(x),\quad\Psi_{12}^+(x);\]

\noindent (2) if
\[{\rm{Re}}\,k\ne0,\qquad C_7=0,\qquad (C_8-C_3C_6+C_2C_7)(C_2C_8-C_4C_6)\ne0\]
or
\[{\rm{Re}}\,k\ne0,\qquad C_2C_8-C_4C_6=0,\qquad (|C_2|+|C_4|)C_7(C_8-C_3C_6+C_2C_7)\ne0\]
then for the eigenvalue $\lambda$ there is the only (up to a constant factor) normalizable vector-eigenfunction $\Psi_{11}^+(x)$ of the Hamiltonian $H_-$;\\

\noindent (3) if
\[{\rm{Re}}\,k\ne0,\qquad C_2=C_4=0,\qquad C_7(C_8-C_3C_6)\ne0\]
then for the eigenvalue $\lambda$ there is the only (up to a constant factor) normalizable vector-eigenfunction $\Psi_{12}^+(x)$ of the Hamiltonian $H_-$;\\

\noindent (4) if
\[C_7=C_2C_8-C_4C_6=0,\qquad C_8-C_3C_6+C_2C_7\ne0\]
or
\[C_7=C_8-C_3C_6+C_2C_7=0,\qquad C_2C_8-C_4C_6\ne0\]
or
\[C_8-C_3C_6+C_2C_7=C_2C_8-C_4C_6=0,\qquad C_7\ne0\]
then for the spectral value $\lambda$ there is no a normalizable vector-eigenfunction of the Hamiltonian $H_-$.

It follows from (\ref{qpr2}) in view of (\ref{vmp2.1}) that the potential of the new Hamiltonian $H_-$ can be reduced with the help of a similarity transformation produced by a constant nondegenerate $2\times2$ matrix either to a diagonal form or to a upper triangular form with equal diagonal elements. Let us consider these situations more detailedly.

If the determinant of the matrix from the last term of ({\ref{qpr2}}) is nonzero,
\begin{equation}4C_6(C_3C_8-C_4C_7)-(C_8+C_3C_6-C_2C_7)^2\ne0,\la{detnz}\end{equation}
then the formulae (\ref{pflrl}) -- (\ref{H71}) and (\ref{psi73}) for $C_6\ne0$ can be simplified as follows,
\begin{align}\tilde\Phi_1^-(x)&={\bf C}^{-1}\bigg\{\Phi_1^-(x)\!+\!{1\over C_6}\bigg[{{2(C_2C_8-C_4C_6)}\over{C_8\!+\!C_2C_7\!-\!C_3C_6\!-\!\Delta}}\!-\!C_2\bigg]\Phi_2^-(x)\bigg\}=\begin{pmatrix}e^{kx}\!+\!\tilde C_2e^{-kx}\\0\end{pmatrix},\nonumber\\
\tilde\Phi_2^-(x)&={\bf C}^{-1}\bigg[{{\Delta\!-\!C_8\!+\!C_2C_7\!+\!C_3C_6}\over{2\Delta}}\Phi_2^-(x)-{{C_6C_7}\over\Delta}\Phi_1^-(x)\bigg]=\begin{pmatrix}0\\C_7e^{kx}\!+\!\tilde C_8e^{-kx}\end{pmatrix},\nonumber\\
\tilde W(x)&=\big[e^{kx}+\tilde C_2e^{-kx}\big]\big[C_7e^{kx}+\tilde C_8e^{-kx}\big]=W(x),\nonumber\\
{\bf C}^{-1}Q_1^\pm{\bf C}&=\mp I_2\partial-k\begin{pmatrix}{{e^{kx}-\tilde C_2e^{-kx}}\over{e^{kx}+\tilde C_2e^{-kx}}}&0\\
0&{{C_7e^{kx}-\tilde C_8e^{-kx}}\over{C_7e^{kx}+\tilde C_8e^{-kx}}}\end{pmatrix},\nonumber\\
{\bf C}^{-1}U_0(x){\bf C}&=-k^2I_2,\nonumber\\
{\bf C}^{-1}H_-{\bf C}&=-I_2\partial^2-\begin{pmatrix}{{8k^2\tilde C_2}\over{[e^{kx}+\tilde C_2e^{-kx}]^2}}&0\\0&{{8k^2C_7\tilde C_8}\over{[C_7e^{kx}+\tilde C_8e^{-kx}]^2}}\end{pmatrix},\la{twoH}\\
\tilde\Psi_1^+(x)&=\!{1\over{2k\tilde C_2}}{\bf C}^{-1}\!\bigg\{\!\Psi_{11}^+(x)\!+\!{1\over C_6}\bigg[{{2(C_2C_8-C_4C_6)}\over{C_8\!+\!C_2C_7\!-\!C_3C_6\!-\!\Delta}}\!-\!C_2\bigg]\Psi_{12}^+(x)\!\bigg\}\!=\!\begin{pmatrix}\!\!{1\over{e^{kx}\!+\tilde C_2e^{\!-kx}}}\!\!\\0\end{pmatrix},\nonumber\\
\tilde\Psi_2^+(x)&=\!{1\over{2kC_7\tilde C_8}}{\bf C}^{-1}\!\bigg[\!{{\Delta\!-\!C_8\!+\!C_2C_7\!+\!C_3C_6}\over{2\Delta}}\Psi_{12}^+(x)\!-\!{{C_6C_7}\over\Delta}\Psi_{11}^+(x)\bigg]\!=\!\begin{pmatrix}0\\\!{1\over{C_7e^{kx}\!+\tilde C_8e^{\!-kx}}}\!\end{pmatrix},\nonumber\\
{\bf C}&=\begin{pmatrix}1&0\\{{C_3C_8-C_4C_7}\over{C_8-C_2C_7}}&1\end{pmatrix},\qquad{\bf C}^{-1}=\begin{pmatrix}1&0\\-{{C_3C_8-C_4C_7}\over{C_8-C_2C_7}}&1\end{pmatrix},\qquad\det{\bf C}=1,\nonumber\\
\tilde C_2&={{2(C_2C_8-C_4C_6)}\over{C_8+C_2C_7-C_3C_6-\Delta}},\qquad\tilde C_8={1\over2}\big[C_8+C_2C_7-C_3C_6-\Delta\big],\nonumber\end{align}

\begin{equation}\Delta=\sqrt{(C_8+C_3C_6-C_2C_7)^2-4C_6(C_3C_8-C_4C_7)},\la{oprDel}\end{equation}
and for $C_6=0$ as follows,
\begin{align}\tilde\Phi_1^-(x)&={\bf C}^{-1}\bigg[\Phi_1^-(x)-{{C_4-C_2C_3}\over{C_8-C_2C_7}}\Phi_2^-(x)\bigg]=\begin{pmatrix}e^{kx}+C_2e^{-kx}\\0\end{pmatrix},\nonumber\\
\tilde\Phi_2^-(x)&={\bf C}^{-1}\Phi_2^-(x)=\begin{pmatrix}0\\C_7e^{kx}+C_8e^{-kx}\end{pmatrix},\nonumber\\
\tilde W(x)&=\big[e^{kx}+C_2e^{-kx}\big]\big[C_7e^{kx}+C_8e^{-kx}\big]=W(x),\nonumber\\
{\bf C}^{-1}Q_1^\pm{\bf C}&=\mp I_2\partial-k\begin{pmatrix}{{e^{kx}-C_2e^{-kx}}\over{e^{kx}+C_2e^{-kx}}}&0\\
0&{{C_7e^{kx}-C_8e^{-kx}}\over{C_7e^{kx}+C_8e^{-kx}}}\end{pmatrix},\nonumber\\
{\bf C}^{-1}U_0(x){\bf C}&=-k^2I_2,\nonumber\\
{\bf C}^{-1}H_-{\bf C}&=-I_2\partial^2-\begin{pmatrix}{{8k^2C_2}\over{[e^{kx}+C_2e^{-kx}]^2}}&0\\0&{{8k^2C_7C_8}\over{[C_7e^{kx}+C_8e^{-kx}]^2}}\end{pmatrix},\la{twoH'}
\end{align}
\begin{align}
\tilde\Psi_1^+(x)&={1\over{2kC_2}}{\bf C}^{-1}\bigg[\Psi_{11}^+(x)-{{C_4-C_2C_3}\over{C_8-C_2C_7}}\Psi_{12}^+(x)\bigg]=\begin{pmatrix}{1\over{e^{kx}+C_2e^{-kx}}}\\0\end{pmatrix},\nonumber\\
\tilde\Psi_2^+(x)&={1\over{2kC_7C_8}}{\bf C}^{-1}\Psi_{12}^+(x)=\begin{pmatrix}0\\{1\over{C_7e^{kx}+C_8e^{-kx}}}\end{pmatrix},\nonumber\\
{\bf C}&=\begin{pmatrix}1&0\\{{C_3C_8-C_4C_7}\over{C_8-C_2C_7}}&1\end{pmatrix},\qquad{\bf C}^{-1}=\begin{pmatrix}1&0\\-{{C_3C_8-C_4C_7}\over{C_8-C_2C_7}}&1\end{pmatrix},\qquad\det{\bf C}=1,\nonumber\end{align}
where $\tilde W(x)$ is the Wronskian of $\tilde\Phi_1^-(x)$ and $\tilde\Phi_2^-(x)$ and the root (\ref{oprDel}) has arbitrary value such that $C_8\!+\!C_2C_7\!-\!C_3C_6\!-\!\Delta\ne0$ (this condition can be satisfied due to (\ref{detnz})). It is evident here and in the what follows below in this Subsection~\ref{example2} that the representations and the intertwinings (\ref{dopsootn75}) transform trivially into the analogous formulae for the Hamiltonians ${\bf C}^{-1}H_+{\bf C}=H_+=-\partial^2$ and ${\bf C}^{-1}H_-{\bf C}$, for the matrix ${\bf C}^{-1}U_0(x){\bf C}$ and for the intertwining operators ${\bf C}^{-1}Q_1^+{\bf C}$ and ${\bf C}^{-1}Q_1^-{\bf C}$, that $\tilde \Psi_{1}^+(x)$ and $\tilde\Psi_{2}^+(x)$ are vector-eigenfunctions (formal sometimes) of the Hamiltonian ${\bf C}^{-1}H_-{\bf C}$ for the same eigenvalue $\lambda=-k^2$ and that $\tilde\Phi_1^-(x)$ and $\tilde\Phi_2^-(x)$ are transformation vector-functions corresponding to conversion of the Hamiltonian ${\bf C}^{-1}H_+{\bf C}=H_+$ to the Hamiltonian ${\bf C}^{-1}H_-{\bf C}$  with the help of the intertwining operator ${\bf C}^{-1}Q_1^-{\bf C}$. It follows from (\ref{twoH}) and (\ref{twoH'}) that any of two diagonal elements of the potential of the reduced Hamiltonian (\ref{twoH}) or (\ref{twoH'}) is ether zero or the potential of P\"oschl -- Teller. 

For
\begin{equation}4C_6(C_3C_8-C_4C_7)-(C_8+C_3C_6-C_2C_7)^2=0,\qquad |C_6|+|C_3C_8-C_4C_7|\ne0\la{cond85}\end{equation}
the formulae (\ref{pflrl}) -- (\ref{H71}) and (\ref{psi73}) convert into the following,
\begin{align}\tilde \Phi_1^-(x)&={\bf C}^{-1}\bigg[\bigg(\sqrt{C_6}+{{\sqrt{C_7}\sqrt{C_2C_8-C_4C_6}\sqrt{C_3C_8-C_4C_7}^*}\over{|C_6|+|C_3C_8-C_4C_7|}}\bigg)\Phi_1^-(x)\nonumber\\
&-{{C_2\sqrt{C_6}^*+C_4\sqrt{C_3C_8-C_4C_7}^*}\over{|C_6|+|C_3C_8-C_4C_7|}}\Phi_2^-(x)\bigg]=\begin{pmatrix}e^{kx}\\-\sqrt{-\tilde C_4}\sqrt{\tilde C_7}e^{kx}+\tilde C_4e^{-kx}\end{pmatrix},
\nonumber\\
\tilde\Phi_2^-(x)&={\bf C}^{-1}\bigg[{{\sqrt{C_6}^*+C_3\sqrt{C_3C_8-C_4C_7}^*}\over{|C_6|+|C_3C_8-C_4C_7|}}\Phi_2^-(x)-
{{C_7\sqrt{C_3C_8-C_4C_7}^*}\over{|C_6|+|C_3C_8-C_4C_7|}}\Phi_1^-(x)\bigg]\nonumber\\
&=
\begin{pmatrix}e^{-kx}\\\tilde C_7e^{kx}+\sqrt{-\tilde C_4}\sqrt{\tilde C_7}\,e^{-kx}\end{pmatrix},\nonumber\\
\tilde W(x)&=\Big[\sqrt{\tilde C_7}e^{kx}+\sqrt{-\tilde C_4}\,e^{-kx}\Big]^2={1\over\alpha}W(x),\nonumber\\
{\bf C}^{-1}Q_1^\pm{\bf C}&=
\mp I_2\partial-k\begin{pmatrix}{{\sqrt{\tilde C_7}e^{kx}-\sqrt{-\tilde C_4}\,e^{-kx}}\over
{\sqrt{\tilde C_7}e^{kx}+\sqrt{-\tilde C_4}\,e^{-kx}}}&-{{2}\over
{\big[\sqrt{\tilde C_7}e^{kx}+\sqrt{-\tilde C_4}\,e^{-kx}\big]^2}}\\
0&{{\sqrt{\tilde C_7}e^{kx}-\sqrt{-\tilde C_4}\,e^{-kx}}\over
{\sqrt{\tilde C_7}e^{kx}+\sqrt{-\tilde C_4}\,e^{-kx}}}\end{pmatrix},\nonumber\\
{\bf C}^{-1}U_0(x){\bf C}&=-k^2I_2,\nonumber\\
{\bf C}^{-1}H_-{\bf C}&=
-I_2\partial^2-8k^2\begin{pmatrix}{{\sqrt{\tilde C_7}\sqrt{-\tilde C_4}}\over
{\big[\sqrt{\tilde C_7}e^{kx}+\sqrt{-\tilde C_4}\,e^{-kx}\big]^2}}&{{\sqrt{\tilde C_7}e^{kx}-\sqrt{-\tilde C_4}\,e^{-kx}}\over
{\big[\sqrt{\tilde C_7}e^{kx}+\sqrt{-\tilde C_4}\,e^{-kx}\big]^3}}\\
0&{{\sqrt{\tilde C_7}\sqrt{-\tilde C_4}}\over
{\big[\sqrt{\tilde C_7}e^{kx}+\sqrt{-\tilde C_4}\,e^{-kx}\big]^2}}\end{pmatrix},\nonumber
\end{align}
\begin{align}
\tilde\Psi_1^+(x)&={\sqrt{\alpha}\over{2k\sqrt{C_7}\sqrt{C_2C_8-C_4C_6}}}\,{\bf C}^{-1}\bigg[\sqrt{C_6}\sqrt{C_7}\Psi_{11}^+(x)+\sqrt{C_2C_3-C_4}\Psi_{12}^+(x)\bigg]\nonumber\\
&=\begin{pmatrix}{1\over{\sqrt{\tilde C_7}e^{kx}+\sqrt{-\tilde C_4}e^{-kx}}}\\
0\end{pmatrix},\nonumber\\
\tilde\Psi_2^+(x)&={\sqrt{\alpha}\over{2k\sqrt{C_7}\sqrt{C_2C_8-C_4C_6}}}\nonumber\\
&\times{\bf C}^{-1}\bigg[\bigg(\sqrt{C_2C_3-C_4}+2\sqrt{C_7}{{C_2\sqrt{C_6}^*+C_4\sqrt{C_3C_8-C_4C_7}^*}\over{|C_6|+|C_3C_8-C_4C_7|}}\bigg)\Psi_{12}^+(x)\nonumber\\
&-\!\!\bigg(\!\!\sqrt{C_6}\!+\!2\sqrt{C_7}{{\sqrt{C_2C_8\!\!-\!C_4C_6}\sqrt{C_3C_8\!\!-\!C_4C_7}^*}\over{|C_6|+|C_3C_8-C_4C_7|}}\!\bigg)\!\Psi_{11}^+(x)\!\bigg]\!\!=\!\!\begin{pmatrix}\!\!{{\sqrt{\tilde C_7}e^{kx}-\sqrt{-\tilde C_4}e^{\!-kx}}\over{\!\big[\!\sqrt{\tilde C_7}e^{kx}\!+\!\sqrt{-\tilde C_4}e^{\!-kx}\!\big]^{\!2}\!}}\!\!\\\!
{{2\sqrt{\tilde C_7}\sqrt{-\tilde C_4}}\over{\sqrt{\tilde C_7}e^{kx}+\sqrt{-\tilde C_4}e^{\!-kx}}}\!
\end{pmatrix}\!,\nonumber
\end{align}
\[{\bf C}\!=\!\begin{pmatrix}\sqrt{C_6}&\!-{{\alpha\sqrt{C_3C_8-C_4C_7}^*}\over{|C_6|+|C_3C_8-C_4C_7|}}\!\\
\!\sqrt{C_3C_8\!-\!C_4C_7}\!&{{\alpha\sqrt{C_6}^*}\over{|C_6|+|C_3C_8-C_4C_7|}}\end{pmatrix},\quad
{\bf C}^{-1}\!=\!\begin{pmatrix}\!{{\sqrt{C_6}^*}\over{|C_6|+|C_3C_8-C_4C_7|}}&{{\sqrt{C_3C_8-C_4C_7}^*}\over{|C_6|+|C_3C_8-C_4C_7|}}\!\\
-{\sqrt{C_3C_8-C_4C_7}\over\alpha}&{\sqrt{C_6}\over\alpha}\end{pmatrix}\!,\]
\[\det{\bf C}=\alpha,\qquad \alpha\in{\Bbb C},\quad\alpha\ne0,\]
\[\tilde C_4=-{1\over\alpha}(C_2C_8-C_4C_6), \qquad\tilde C_7={1\over\alpha}C_7,\]
where $*$ denotes complex conjugation, the roots $\sqrt{C_6}$, $\sqrt{C_7}$, $\sqrt{C_2C_8\!-\!C_4C_6}$, $\sqrt{C_3C_8\!-\!C_4C_7}$, $\sqrt{C_2C_3-C_4}$, $\sqrt{\tilde C_7}$, $\sqrt{-\tilde C_4}$ and $\sqrt\alpha$ have arbitrary values satisfying the following conditions,
\[\sqrt{C_6}\sqrt{C_3C_8-C_4C_7}={1\over2}\big[C_8+C_3C_6-C_2C_7\big],\]
\[\sqrt{C_7}\sqrt{C_2C_8-C_4C_6}={1\over2}\big[C_8-C_3C_6+C_2C_7\big],\]
\[\sqrt{\tilde C_7}={\sqrt{C_7}\over\sqrt\alpha},\qquad\sqrt{-\tilde C_4}={\sqrt{C_2C_8-C_4C_6}\over\sqrt\alpha},\]
\begin{equation}C_8=C_2C_7+C_3C_6+2\sqrt{C_6}\sqrt{C_7}\sqrt{C_2C_3-C_4},\la{c889}\end{equation}
the roots $\sqrt{C_2C_8-C_4C_6}$ and $\sqrt{C_3C_8-C_4C_7}$ as well as $C_8$ (see (\ref{c889})) can be expressed through the constants $C_2$, $C_3$, $C_4$, $C_6$ and $C_7$,
\[\sqrt{C_2C_8-C_4C_6}=C_2\sqrt{C_7}+\sqrt{C_6}\sqrt{C_2C_3-C_4},\]
\[\sqrt{C_3C_8-C_4C_7}=C_3\sqrt{C_6}+\sqrt{C_7}\sqrt{C_2C_3-C_4}\,\]
and $\tilde W(x)$ is the Wronskian of $\tilde\Phi_1^-(x)$ and $\tilde\Phi_2^-(x)$. The possibility to define the roots $\sqrt{C_6}$, $\sqrt{C_7}$, $\sqrt{C_2C_8-C_4C_6}$, $\sqrt{C_3C_8-C_4C_7}$ and $\sqrt{C_2C_3-C_4}$ so that the relations (\ref{c889}) hold is provided by the first of the conditions~(\ref{cond85}).

At last, if
\[4C_6(C_3C_8-C_4C_7)-(C_8+C_3C_6-C_2C_7)^2=0,\qquad C_6=C_3C_8-C_4C_7=0\]
then
\begin{align}
W(x)&=C_7[e^{kx}+C_2e^{-kx}]^2\quad\Rightarrow\quad C_7\ne0,\nonumber\\
Q_1^\pm&=\mp I_2\partial-k{{e^{kx}-C_2e^{-kx}}\over{e^{kx}+C_2e^{-kx}}}I_2,\nonumber
\end{align}
\begin{align}
U_0(x)&=-k^2I_2,\nonumber\\
H_-&=-I_2\partial^2-{{8k^2C_2}\over{[e^{kx}+C_2e^{-kx}]^2}}I_2,\la{gam85}
\end{align}
\[\Phi_1^-(x)=\begin{pmatrix}1\\C_3\end{pmatrix}[e^{kx}+C_2e^{-kx}],\qquad
\Phi_2^-(x)=C_7\begin{pmatrix}0\\1\end{pmatrix}[e^{kx}+C_2e^{-kx}],\]
\[\Psi_{11}^+(x)=2kC_2\begin{pmatrix}1\\C_3\end{pmatrix}{1\over{e^{kx}+C_2e^{-kx}}},\qquad
\Psi_{12}^+(x)=2kC_2C_7\begin{pmatrix}0\\1\end{pmatrix}{1\over{e^{kx}+C_2e^{-kx}}}\]
and it is possible to use the vector-functions 
\[\tilde\Phi_1^-(x)\!=\!\Phi_1^-(x)\!-\!{C_3\over C_7}\Phi_2^-(x)\!=\!\begin{pmatrix}e^{kx}\!+\!C_2e^{-kx}\\0\end{pmatrix},\qquad\tilde\Phi_2^-(x)\!=\!{1\over C_7}\Phi_2^-(x)\!=\!\begin{pmatrix}0\\e^{kx}\!+\!C_2e^{-kx}\end{pmatrix}\]
as transformation vector-functions instead of $\Phi_1^-(x)$ and $\Phi_2^-(x)$ and the vector-functions
\[\tilde\Psi_1^+(x)\!=\!{1\over{2kC_2}}\Big[\Psi_{11}^+(x)\!-\!{C_3\over C_7}\Psi_{12}^+(x)\Big]\!=\!\begin{pmatrix}\!\!{1\over{\!e^{kx}\!+\!C_2e^{\!-kx}\!}}\!\!\\0\end{pmatrix}\!,\,\,\tilde\Psi_2^+(x)\!=\!{1\over{2kC_2C_7}}\Psi_{12}^+(x)\!=\!\begin{pmatrix}0\\{1\over{\!e^{kx}\!+\!C_2e^{\!-kx}\!}}\!\end{pmatrix}\]
as vector-eigenfunctions (formal for $C_2=0$ and normalizable for $C_2\ne0$) instead of $\Psi_{11}^+(x)$ and $\Psi_{12}^+(x)$.
One can see that both diagonal elements of the potential of the new Hamiltonian (\ref{gam85}) are either zeroes for $C_2=0$ or the identical potentials of P\"oschl -- Teller for $C_2\ne0$.

\subsection{Subcase $\lambda_1=\lambda_2$, $g^-_1=1$: adding up to two bound states described by eigen- and associated vector-functions with the same energy value\la{example3}}

In this subcase general form of transformation vector-functions $\Phi_1^-(x)$ and $\Phi_2^-(x)$ is the following in view of (\ref{sobprisfun}),
\[\Phi_1^-(x)\!=\!\begin{pmatrix}-C_1{x\over{2k}}e^{kx}\!+\!C_2{x\over{2k}}e^{-kx}\!+\!C_5e^{kx}\!+\!C_6e^{-kx}\\
-C_3{x\over{2k}}e^{kx}\!+\!C_4{x\over{2k}}e^{-kx}\!+\!C_7e^{kx}\!+\!C_8e^{-kx}\end{pmatrix},\qquad \Phi_2^-(x)\!=\!\begin{pmatrix}C_1e^{kx}\!+\!C_2e^{-kx}\\C_3e^{kx}\!+\!C_4e^{-kx}\end{pmatrix},\]
\begin{equation}H_+\Phi_1^-=\lambda\Phi_1^-+\Phi_2^-,\quad H_+\Phi_2^-=\lambda\Phi_2^-,\qquad\lambda=-k^2\ne0,\la{53pflnl}\end{equation}
where $C_1$, \dots, $C_8$ are arbitrary complex, in general, constants and we assume without the loss of generality that $C_1=1$ and $C_5=0$ (the latter condition can be achieved in any case by the change of a canonical basis in the kernel of $Q_1^-$: $\Phi_1^-(x)-C_5\Phi_2^-(x)$ and $\Phi_2^-(x)$ instead of $\Phi_1^-(x)$ and $\Phi_2^-(x)$). The remaining constants $C_2$, $C_3$, $C_4$, $C_6$, $C_7$ and $C_8$ are chosen so that the Wronskian $W(x)$ of the vector-functions $\Phi_1^-(x)$ and $\Phi_2^-(x)$,
\begin{equation}W(x)=-C_7e^{2kx}-[C_2C_8-C_4C_6]e^{-2kx}-{1\over k}[C_4-C_2C_3]x-[C_8+C_2C_7-C_3C_6],\end{equation}
does not vanish on the real axis. The operators $Q_1^-$ and $Q_1^+$, the matrix $U_0(x)$ and the new Hamiltonian $H_-$ take the following form,
\begin{align}
Q_1^\pm&=\mp I_2\partial+{1\over{W(x)}}\bigg\{
\Big[kC_7e^{2kx}-k\Delta_{28}e^{-2kx}+{1\over{2k}}\Delta_1\Big]I_2\nonumber\\
&\qquad\qquad\qquad\qquad\qquad\qquad+{1\over{2k}}M_1e^{2kx}-{1\over{2k}}{ M}_2e^{-2kx}+{M}_3x+k{M}_4
\bigg\},
\end{align}
\begin{align}
U_0(x)&=-k^2I_2\!+\!{1\over{W(x)}}\!\begin{pmatrix}[e^{kx}\!+\!C_2e^{-kx}][C_3e^{kx}\!+\!C_4e^{-kx}]\!\!&-[e^{kx}+C_2e^{-kx}]^2\\
[C_3e^{kx}+C_4e^{-kx}]^2&\!\!-[e^{kx}\!+\!C_2e^{-kx}][C_3e^{kx}\!+\!C_4e^{-kx}]\end{pmatrix}\nonumber\\
&\equiv-k^2I_2+{1\over{W(x)}}\Big[{ M}_1e^{2kx}+{ M}_2e^{-2kx}+{ M}_3\Big],\\
H_-&=-I_2\partial^2+{2\over{W^2(x)}}\bigg\{
\Big[-2k[\Delta_1x+k(C_8+\Delta_{27})][C_7e^{2kx}+\Delta_{28}e^{-2kx}]\nonumber\\
&\qquad\qquad\qquad\qquad\qquad\qquad+2\Delta_1[C_7e^{2kx}-\Delta_{28}e^{-2kx}]
-8k^2C_7\Delta_{28}+{1\over{2k^2}}\Delta_1^2\Big]I_2\nonumber\\
&-\Big[{1\over k}\Delta_1xe^{2kx}+[C_8+\Delta_{27}-{1\over{2k^2}}\Delta_1]e^{2kx}+4\Delta_{28}\Big]{ M}_1\nonumber\\
&-\Big[{1\over k}\Delta_1xe^{-2kx}+[C_8+\Delta_{27}+{1\over{2k^2}}\Delta_1]e^{-2kx}+4C_7\Big]{ M}_2\nonumber\\
&+\Big[2kx[C_7e^{2kx}\!\!-\!\Delta_{28}e^{-2kx}]\!-\![C_7e^{2kx}\!\!+\!\Delta_{28}e^{-2kx}]\Big]{ M}_3\!+\!2k^2\Big[C_7e^{2kx}\!\!-\!\Delta_{28}e^{-2kx}\Big]{ M}_4
\bigg\},\la{53h100}
\end{align}
\[{ M}_1=\begin{pmatrix}C_3&-1\\C_3^2&-C_3\end{pmatrix},\qquad
{M}_2=\begin{pmatrix}C_2C_4&-C_2^2\\C_4^2&-C_2C_4\end{pmatrix},\]
\[{ M}_3=\begin{pmatrix}C_4+C_2C_3&-2C_2\\2C_3C_4&-C_4-C_2C_3\end{pmatrix},\qquad
{ M}_4=\begin{pmatrix}C_8-\Delta_{27}&-2C_6\\2\Delta_{38}&-[C_8-\Delta_{27}]\end{pmatrix},\]
\[\Delta_1=C_4-C_2C_3,\quad\Delta_{27}=C_2C_7-C_3C_6,\quad\Delta_{28}=C_2C_8-C_4C_6,\quad\Delta_{38}=C_3C_8-C_4C_7,\]
\[2\Delta_{28}{ M}_1+2C_7{ M}_2-(C_8+\Delta_{27}){ M}_3+\Delta_1{ M}_4=0,\]
so that
\begin{equation}H_+=Q_1^+Q_1^-+U_0(x),\qquad H_-=Q_1^-Q_1^++U_0(x),\qquad Q_1^-H_+=H_-Q_1^-.\la{pres99}\end{equation}

For the spectral value $\lambda$ of the Hamiltonian $H_-$ one can easily construct formal vector-eigenfunctions and formal associated vector-functions of the first order
\begin{align}
\Psi_{1,0}^+(x)&=Q_1^-\begin{pmatrix}e^{kx}\\0\end{pmatrix}={1\over{2kW(x)}}
\bigg[C_3\begin{pmatrix}1\\C_3\end{pmatrix}e^{3kx}
+4kC_3\begin{pmatrix}C_2\\C_4\end{pmatrix}xe^{kx}-\begin{pmatrix}4k^2\Delta_{27}-\Delta_1\\
-4k^2\Delta_{38}\end{pmatrix}e^{kx}\nonumber\\
&\qquad\qquad\qquad\quad-\begin{pmatrix}4k^2\Delta_{28}+
C_2C_4\\C_4^2\end{pmatrix}e^{-kx}\bigg],\nonumber\\
\Psi_{1,1}^+(x)&=Q_1^-\begin{pmatrix}-{xe^{kx}\over{2k}}\\0\end{pmatrix}\!=\!{1\over{4k^2W(x)}}
\bigg[\!-\!\begin{pmatrix}C_3x\!-\!2kC_7\\C_3^2x\end{pmatrix}\!e^{3kx}
\!\!+\!\!\begin{pmatrix}(4k^2\Delta_{28}\!+\!C_2C_4)x\!+\!2k\Delta_{28}\\C_4^2x\end{pmatrix}\!e^{-kx}\nonumber\\
&\qquad\qquad\qquad\quad-\begin{pmatrix}4kC_2C_3x^2-(\Delta_1+4k^2\Delta_{27})x-2k(C_8+\Delta_{27})\\
4kC_3C_4x^2+4k^2\Delta_{38}x\end{pmatrix}e^{kx}\bigg],\nonumber\\
\Psi_{2,0}^+(x)&=Q_1^-\begin{pmatrix}e^{-kx}\\0\end{pmatrix}={1\over{2kW(x)}}
\bigg[\!\!-\!C_4\begin{pmatrix}C_2\\C_4\end{pmatrix}e^{-3kx}
\!\!+\!4kC_4\begin{pmatrix}1\\C_3\end{pmatrix}xe^{-kx}\!\!+\!\!\begin{pmatrix}4k^2C_8\!+\!\Delta_1\\4k^2\Delta_{38}\end{pmatrix}e^{-kx}\nonumber\\
&\qquad\qquad\qquad\quad+\begin{pmatrix}4k^2C_7+C_3\\C_3^2\end{pmatrix}e^{kx}\bigg],\nonumber\\
\Psi_{2,1}^+(x)&=Q_1^-\begin{pmatrix}{xe^{-kx}\over{2k}}\\0\end{pmatrix}\!=\!{1\over{4k^2W(x)}}
\bigg[\!-\!\begin{pmatrix}C_2C_4x\!+\!2k\Delta_{28}\\C_4^2x\end{pmatrix}\!e^{-3x}
\!+\!\begin{pmatrix}(4k^2C_7\!+\!C_3)x\!-\!2kC_7\\C_3^2x\end{pmatrix}\!e^{kx}\nonumber\\
&\qquad\qquad\qquad\quad+\begin{pmatrix}4kC_4x^2-(\Delta_1-4k^2C_8)x-2k(C_8+\Delta_{27})\\4kC_3C_4x^2+4k^2\Delta_{38}x\end{pmatrix}e^{-kx}\bigg],\nonumber
\end{align}
\begin{align}
\Psi_{3,0}^+(x)&=Q_1^-\begin{pmatrix}0\\e^{kx}\end{pmatrix}={1\over{2kW(x)}}
\bigg[-\begin{pmatrix}1\\C_3\end{pmatrix}e^{3kx}
-4k\begin{pmatrix}C_2\\C_4\end{pmatrix}xe^{kx}-\begin{pmatrix}4k^2C_6\\4k^2C_8-\Delta_1\end{pmatrix}e^{kx}\nonumber\\
&\qquad\qquad\qquad\quad+\begin{pmatrix}C_2^2\\-4k^2\Delta_{28}+C_2C_4\end{pmatrix}e^{-kx}\bigg],\nonumber\\
\Psi_{3,1}^+(x)&=Q_1^-\begin{pmatrix}0\\-{xe^{kx}\over{2k}}\end{pmatrix}\!=\!{1\over{4k^2W(x)}}
\bigg[\!\!\begin{pmatrix}x\\C_3x\!+\!2kC_7\end{pmatrix}e^{3kx}
\!\!-\!\begin{pmatrix}C_2^2x\\\!-(4k^2\Delta_{28}\!-\!C_2C_4)x\!-\!2k\Delta_{28}\end{pmatrix}\!e^{-kx}\nonumber\\
&\qquad\qquad\qquad\quad+\begin{pmatrix}4kC_2x^2+4k^2C_6x\\4kC_4x^2+(\Delta_1+4k^2C_8)x+2k(C_8+\Delta_{27})\end{pmatrix}e^{kx}\bigg],\nonumber\\
\Psi_{4,0}^+(x)&=Q_1^-\begin{pmatrix}0\\e^{-kx}\end{pmatrix}\!=\!{1\over{2kW(x)}}
\bigg[C_2\begin{pmatrix}C_2\\C_4\end{pmatrix}e^{-3kx}
\!-\!4kC_2\begin{pmatrix}1\\C_3\end{pmatrix}xe^{-kx}\!\!-\!\begin{pmatrix}4k^2C_6\\\!-4k^2\Delta_{27}\!-\!\Delta_1\end{pmatrix}e^{-kx}\nonumber\\
&\qquad\qquad\qquad\quad-\begin{pmatrix}1\\-4k^2C_7+C_3\end{pmatrix}e^{kx}\bigg],\nonumber\\
\Psi_{4,1}^+(x)&=Q_1^-\begin{pmatrix}0\\{xe^{-kx}\over{2k}}\end{pmatrix}\!=\!{1\over{4k^2W(x)}}
\bigg[\!\begin{pmatrix}C_2^2x\\C_2C_4x\!-\!2k\Delta_{28}\end{pmatrix}\!e^{-3kx}
\!-\!\begin{pmatrix}x\\\!-(4k^2C_7\!-\!C_3)x\!+\!2kC_7\end{pmatrix}\!e^{kx}\nonumber\\
&\qquad\qquad\qquad\quad-\begin{pmatrix}4kC_2x^2+4k^2C_6x\\4kC_2C_3x^2+(\Delta_1-4k^2\Delta_{27})x+2k(C_8+\Delta_{27})\end{pmatrix}e^{-kx}\bigg],\nonumber
\end{align}
\begin{equation}H_-\Psi_{i,0}^+=\lambda\Psi_{i,0}^+,\qquad
(H_--\lambda I_2)\Psi_{i,1}^+=\Psi_{i,0}^+,\qquad\qquad i=1,2,3,4,\la{538sobvech}\end{equation}
only six of which are linearly independent in view of the fact that the vector-functions $\Phi_1^-(x)$ and $\Phi_2^-(x)$ (see (\ref{53pflnl})) form a canonical basis in the kernel of $Q_1^-$. The latter leads to the relations
\begin{align}\Psi_{1,0}^+(x)+C_2\Psi_{2,0}^+(x)+C_3\Psi_{3,0}^+(x)+C_4\Psi_{4,0}^+(x)&=0,\nonumber\\ \Psi_{1,1}^+(x)+C_2\Psi_{2,1}^+(x)+C_3\Psi_{3,1}^+(x)+C_4\Psi_{4,1}^+(x)+C_6\Psi_{2,0}^+(x)+C_7\Psi_{3,0}^+(x)+
C_8\Psi_{4,0}^+(x)&=0.\end{align}
It follows from the results of \cite{sokolov13} that in the considered subcase $\lambda_1=\lambda_2$, $g_1^-=1$ there is linear differential operator of the 3-rd order $Q_3^+$ with the coefficient $I_2$ at $\partial^3$ that intertwines the Hamiltonians $H_+$ and $H_-$ in the opposite direction, $Q_3^+H_-=H_+Q_3^+$, and six linearly independent vector-functions from the set (\ref{538sobvech}) form a canonical basis in the kernel of $Q_3^+$ providing an opportunity to construct $Q_3^+$ explicitly with the help of (\ref{qNrep5}).

A linearly independent of (\ref{538sobvech}) formal vector-eigenfunction $\Psi_{5,0}^+(x)$ of the Hamiltonian $H_-$ for the spectral value $\lambda$ can be found in the form
\begin{align}\Psi_{5,0}^+(x)&=Q_1^-\begin{pmatrix}{1\over{8k^2}}(x^2-{x\over k})e^{kx}+{C_2\over{8k^2}}(x^2+{x\over k})e^{-kx}+{C_6\over{2k}}(x+{1\over{2k}})e^{-kx}\\
{C_3\over{8k^2}}(x^2-{x\over k})e^{kx}+{C_4\over{8k^2}}(x^2+{x\over k})e^{-kx}-{C_7\over{2k}}(x-{1\over{2k}})e^{kx}+{C_8\over{2k}}(x+{1\over{2k}})e^{-kx}\end{pmatrix}\nonumber\\
&={1\over{2kW(x)}}\Bigg[C_7\begin{pmatrix}0\\C_7\end{pmatrix}e^{3kx}\!+\!\begin{pmatrix}2C_2C_7-{\Delta_1
\over{2k^2}}\\2C_4C_7\!-\!C_3{\Delta_1\over{2k^2}}\end{pmatrix}x^2e^{kx}\!+\!\begin{pmatrix}4kC_6C_7-{{C_8+\Delta_{27}}
\over{k}}\\4kC_7C_8\!-\!C_3{{C_8+\Delta_{27}}\over{k}}\end{pmatrix}xe^{kx}\nonumber\\
&\quad-C_7\begin{pmatrix}C_6\\-\Delta_{27}\end{pmatrix}e^{kx}-\begin{pmatrix}C_6(C_8+\Delta_{27})\\
C_8^2-C_6\Delta_{38}\end{pmatrix}e^{-kx}+\begin{pmatrix}-C_2{{C_8+\Delta_{27}}
\over{k}}\\4kC_7\Delta_{28}-C_4{{C_8+\Delta_{27}}\over{k}}\end{pmatrix}xe^{-kx}\nonumber\\
&\quad-\begin{pmatrix}2\Delta_{28}+C_2{\Delta_1\over{2k^2}}\\2C_3\Delta_{28}+C_4{\Delta_1\over{2k^2}}\end{pmatrix}x^2e^{-kx}-\Delta_{28}\begin{pmatrix}C_6\\C_8\end{pmatrix}e^{-3kx}\Bigg],
\nonumber\end{align}
\begin{equation}H_-\Psi_{5,0}^+=\lambda\Psi_{5,0}^+,\la{53psi50}\end{equation}
since
\[(H_+\!-\!\lambda I_2)\begin{pmatrix}{1\over{8k^2}}(x^2-{x\over k})e^{kx}+{C_2\over{8k^2}}(x^2+{x\over k})e^{-kx}+{C_6\over{2k}}(x+{1\over{2k}})e^{-kx}\\
{C_3\over{8k^2}}(x^2\!-\!{x\over k})e^{kx}\!+\!{C_4\over{8k^2}}(x^2\!+\!{x\over k})e^{-kx}\!-\!{C_7\over{2k}}(x\!-\!{1\over{2k}})e^{kx}\!+\!{C_8\over{2k}}(x\!+\!{1\over{2k}})e^{-kx}\end{pmatrix}\!=\!\Phi_1^-(x),\]
the vector-function $\Phi_1^-(x)$ belongs to the kernel of $Q_1^-$ and a chain of formal associated vector-functions of the Hamiltonian $H_+$ is mapped (see Subsection \ref{sjio}) by the operator $Q_1^-$ into a chain of formal associated vector-functions of the Hamiltonian $H_-$ (some first terms of the chain can be mapped by $Q_1^-$ into zeroes).

Analysis of the vector-functions (\ref{538sobvech}) and (\ref{53psi50}) leads to the following results:\\

\noindent (1) if
\[{\rm{Re}}\,k\ne0,\qquad C_7(C_2C_8-C_4C_6)\ne0\]
then for the eigenvalue~$\lambda$ of the Hamiltonian $H_-$ there is the only (up to a constant factor) normalizable vector-eigenfunction $\Psi_{6,0}^+(x)$ and the only (up to a constant factor and up to adding of a vector-function proportional to $\Psi_{6,0}^+(x)$) associated vector-function of the first order $\Psi_{6,1}^+(x)$:
\begin{align}\Psi_{6,0}^+(x)&=\Psi_{1,0}^+(x)+C_3\Psi_{3,0}^+(x)=-C_2\Psi_{2,0}^+(x)-C_4\Psi_{4,0}^+(x)
\nonumber\\
&={{2k}\over{W(x)}}\bigg[\begin{pmatrix}{\Delta_1\over{4k^2}}-C_2C_7\\C_3{\Delta_1\over{4k^2}}-C_4C_7\end{pmatrix}e^{kx}-
\begin{pmatrix}C_2{\Delta_1\over{4k^2}}+\Delta_{28}\\C_4{\Delta_1\over{4k^2}}+C_3\Delta_{28}\end{pmatrix}e^{-kx}\bigg],\nonumber\\
\Psi_{6,1}^+(x)&=\Psi_{1,1}^+(x)+C_3\Psi_{3,1}^+(x)+C_7\Psi_{3,0}^+(x)
\nonumber\\
&=-C_2\Psi_{2,1}^+(x)-C_4\Psi_{4,1}^+(x)-C_6\Psi_{2,0}^+(x)-C_8\Psi_{4,0}^+(x)
\nonumber\\
&={{1}\over{W(x)}}\Bigg[\begin{pmatrix}{{\Delta_1}\over{4k^2}}-C_2C_7\\
C_3{{\Delta_1}\over{4k^2}}-C_4C_7\end{pmatrix}xe^{kx}
+\begin{pmatrix}{{C_8+\Delta_{27}}\over{2k}}-2kC_6C_7\\
{{C_3C_8+C_4C_7-C_3^2C_6}\over{2k}}-2kC_7C_8\end{pmatrix}e^{kx}\nonumber\\
&\qquad\qquad+\begin{pmatrix}{{\Delta_{28}+C_2^2C_7}\over{2k}}\\
{{C_3\Delta_{28}+C_2C_4C_7}\over{2k}}-2kC_7\Delta_{28}\end{pmatrix}e^{-kx}+
\begin{pmatrix}C_2{{\Delta_1}\over{4k^2}}+\Delta_{28}\\
C_4{{\Delta_1}\over{4k^2}}+C_3\Delta_{28}\end{pmatrix}xe^{-kx}\Bigg],\nonumber
\end{align}
\begin{equation}H_-\Psi_{6,0}^+=\lambda\Psi_{6,0}^+,\qquad (H_--\lambda I_2)\Psi_{6,1}^+=\Psi_{6,0}^+,\qquad \Psi_{6,0}^+(x),
\Psi_{6,1}^+(x)\in\ker Q_3^+;\la{53psi6061}\end{equation}

\noindent (2) if 
\[{\rm{Re}}\,k\ne0,\qquad C_7=C_4-C_2C_3=0,\qquad (C_8+C_2C_7-C_3C_6)(C_2C_8-C_4C_6)\ne0\]
or
\[{\rm{Re}}\,k\ne0,\qquad C_4-C_2C_3=C_2C_8-C_4C_6=0,\qquad C_2C_7(C_8+C_2C_7-C_3C_6)\ne0\]
or
\[{\rm{Re}}\,k=0,\qquad C_4-C_2C_3\ne0\]
then for the eigenvalue $\lambda$ of the Hamiltonian $H_-$ there is the only (up to a constant factor) normalizable vector-eigenfunction $\Psi_{6,0}^-(x)$ and there is no a normalizable associated vector-function of the first order;\\

\noindent (3) if 
\[{\rm{Re}}\,k\ne0,\qquad C_2=C_4=0,\qquad C_7(C_8-C_3C_6)\ne0\]
then for the eigenvalue $\lambda$ of the Hamiltonian $H_-$ there is the only (up to a constant factor) normalizable vector-eigenfunction \[\Psi_{6,1}^+(x)\Big|_{C_2=C_4=0}=\begin{pmatrix}2kC_6C_7-{{C_8-C_3C_6}\over{2k}}\\2kC_7C_8
-C_3{{C_8-C_3C_6}\over{2k}}\end{pmatrix}{{1}\over{C_7e^{kx}\!+\!(C_8\!-\!C_3C_6)e^{-kx}}},\]
\[H_-\Psi_{6,1}^+\Big|_{C_2=C_4=0}=\lambda\Psi_{6,1}^+\Big|_{C_2=C_4=0}\]
(cf. with (\ref{53psi6061})) and there is no a normalizable associated vector-function of the first order;\\

\noindent (4) if
\[{\rm{Re}}\,k\ne0,\qquad C_7=0,\qquad(C_4-C_2C_3)(C_2C_8-C_4C_6)\ne0\]
or
\[{\rm{Re}}\,k\ne0,\qquad C_2C_8-C_4C_6=0,\qquad C_7(C_4-C_2C_3)\ne0\]
or
\[{\rm{Re}}\,k\ne0,\qquad C_7=C_2C_8-C_4C_6=0\]
or
\[{\rm{Re}}\,k=0,\qquad C_4-C_2C_3=0\]
then for the eigenvalue $\lambda$ of the Hamiltonian $H_-$ there is no a normalizable vector-eigenfunction.

For $\Delta_1\ne0\Leftrightarrow C_4\ne C_2C_3$ the formulae (\ref{53pflnl}) -- (\ref{53h100}) and (\ref{53psi6061}) can be simplified with the help of similarity transformation as follows,
\begin{align}
\tilde\Phi_1^-(x)&={\bf C}^{-1}\Big[\Phi_1^-(x)+{{C_2C_7}\over\Delta_1}\Phi_2^-(x)\Big]=\begin{pmatrix}-{x\over{2k}}e^{kx}
+\tilde C_6e^{-kx}\\{x\over{2k}}e^{-kx}+\tilde C_7e^{kx}+\tilde C_8e^{-kx}\end{pmatrix},\nonumber\\
\tilde\Phi_2^-(x)&={\bf C}^{-1}\Phi_2^-(x)=\begin{pmatrix}e^{kx}\\e^{-kx}\end{pmatrix},\nonumber\\
\tilde W(x)&=-\tilde C_7e^{2kx}+\tilde C_6e^{-2kx}-{1\over k}[x+k\tilde C_8]={1\over\Delta_1}W(x),\nonumber\\
{\bf C}^{-1}Q_1^\pm{\bf C}&=\!\mp I_2\partial\!+\!{k\over{\!\tilde W(x)\!}}\Bigg\{\!
\Big[\tilde C_7e^{2kx}\!+\!\tilde C_6e^{\!-2kx}\!+\!{1\over{\!2k^2\!}}\Big]\!I_2\!+\!2\!\begin{pmatrix}{1\over{2k}}[x+k\tilde C_8]&\!-[{{\!e^{2kx}\!}\over{\!4k^2\!}}\!+\!\tilde C_6]\!\\\!-[\tilde C_7\!+\!{e^{\!-2kx}\over{4k^2}}]\!&\!-{1\over{2k}}[x\!+\!k\tilde C_8]\!\end{pmatrix}
\!\!\Bigg\},\nonumber\\
{\bf C}^{-1}U_0(x){\bf C}&=-k^2I_2+{1\over{\tilde W(x)}}\begin{pmatrix}1&-e^{2kx}\\
e^{-2kx}&-1\end{pmatrix},\nonumber\\
{\bf C}^{-1}H_-{\bf C}&=\!-I_2\partial^2\!\!+\!{{8k}\over{\!\tilde W^2(x)\!\!}}\Bigg\{\!
[x\!+\!k\tilde C_8]\!\!\begin{pmatrix}\!\!\!\tilde C_6e^{\!-2kx}\!\!\!\!&{e^{2kx}\over{4k^2}}\\
-{{e^{\!-2kx}\!}\over{4k^2}}&\!\!-\tilde C_7e^{2kx}\!\!\end{pmatrix}\!\!-\!k[\tilde C_7e^{2kx}\!\!\!-\!\tilde C_6e^{\!-2kx}]\!\begin{pmatrix}{1\over{4k^2}}&\tilde C_6\\
\!\!-\tilde C_7\!\!&\!\!-{1\over{\!4k^2}}\!\end{pmatrix}\nonumber\\
&+2k\!\begin{pmatrix}\![\tilde C_7e^{kx}\!+\!{e^{-kx}\over{4k^2}}][{{e^{kx}}\over{4k^2}}+\tilde C_6e^{-kx}]\!&-[{e^{kx}\over{4k^2}}+\tilde C_6e^{-kx}]^2\\
-[\tilde C_7e^{kx}+{e^{-kx}\over{4k^2}}]^2&\![\tilde C_7e^{kx}\!+\!{e^{-kx}\over{4k^2}}][{{e^{kx}\!}\over{4k^2}}+\tilde C_6e^{-kx}]\end{pmatrix}\!\!
\Bigg\},\nonumber\\
\tilde\Psi_{1,0}^+(x)&={\bf C}^{-1}\Psi_{6,0}^+(x)={{2k}\over{\tilde W(x)}}\begin{pmatrix}{1\over{4k^2}}e^{kx}+\tilde C_6e^{-kx}\\
-[\tilde C_7e^{kx}+{1\over{4k^2}}e^{-kx}]\end{pmatrix},\nonumber\\
\tilde\Psi_{1,1}^+(x)&={\bf C}^{-1}\Big[\Psi_{6,1}^+(x)-{{C_8-C_2C_7-C_3C_6}\over{2\Delta_1}}\Psi_{6,0}^+(x)\Big]
\nonumber\\
&={{2k}\over{\tilde W(x)}}\begin{pmatrix}{{x+k\tilde C_8}\over{2k}}[{e^{kx}\over{4k^2}}-\tilde C_6e^{-kx}]-\tilde C_6[\tilde C_7e^{kx}+{e^{-kx}\over{4k^2}}]\\
\tilde C_7[{e^{kx}\over{4k^2}}+\tilde C_6e^{-kx}]
-{{x+k\tilde C_8}\over{2k}}[\tilde C_7e^{kx}-{{e^{-kx}}\over{4k^2}}]
\end{pmatrix},\nonumber
\end{align}

\[{\bf C}=\begin{pmatrix}1&C_2\\C_3&C_4\end{pmatrix},\qquad{\bf C}^{-1}={1\over\Delta_1}\begin{pmatrix}C_4&-C_2\\-C_3&1\end{pmatrix},\qquad\det{\bf C}=\Delta_1,\]
\[\tilde C_6=-{\Delta_{28}\over\Delta_1},\qquad\tilde C_7={C_7\over\Delta_1},\qquad\tilde C_8={{C_8+\Delta_{27}}\over\Delta_1},\]
and if $\Delta_1=0\Leftrightarrow C_4=C_2C_3$ then the formulae (\ref{53pflnl}) -- (\ref{53h100}) and (\ref{53psi6061}) can be simplified as well,
\begin{align}
\tilde\Phi_1^-(x)&={\bf C}^{-1}\Phi_1^-(x)=\begin{pmatrix}-{x\over{2k}}e^{kx}+C_2{x\over{2k}}e^{-kx}+C_6e^{-kx}\\
\tilde C_7e^{kx}+\tilde C_8e^{-kx}\end{pmatrix},\nonumber\\
\tilde\Phi_2^-(x)&={\bf C}^{-1}\Phi_2^-(x)=\begin{pmatrix}e^{kx}+C_2e^{-kx}\\0\end{pmatrix},\nonumber\\
\tilde W(x)&=-\Big[e^{kx}+C_2e^{-kx}\Big]\Big[\tilde C_7e^{kx}+\tilde C_8e^{-kx}\Big]=-{1\over\alpha}W(x),\nonumber\\
{\bf C}^{-1}Q_1^\pm{\bf C}
&=\mp I_2\partial-\begin{pmatrix}k{{e^{kx}-C_2e^{-kx}}\over{e^{kx}+C_2e^{-kx}}}&-{1\over{2k}}{{e^{2kx}-C_2^2e^{-2kx}+4k(C_2x+kC_6)}\over{[e^{kx}+C_2e^{-kx}][\tilde C_7e^{kx}+\tilde C_8e^{-kx}]}}\\0&
k{{\tilde C_7e^{kx}-\tilde C_8e^{-kx}}\over{\tilde C_7e^{kx}+\tilde C_8e^{-kx}}}\end{pmatrix},\nonumber\\
{\bf C}^{-1}U_0(x){\bf C}
&=\begin{pmatrix}-k^2&{{e^{kx}+C_2e^{-kx}}\over{\tilde C_7e^{kx}+\tilde C_8e^{-kx}}}\\0&-k^2\end{pmatrix},\nonumber\\
{\bf C}^{-1}H_-{\bf C}&=-I_2\partial^2-\begin{pmatrix}{{8k^2C_2}\over{[e^{kx}+C_2e^{-kx}]^2}}&8k{{[C_2x+kC_6][\tilde C_7e^{2kx}-C_2\tilde C_8e^{-2kx}]}\over{[e^{kx}+C_2e^{-kx}]^2[\tilde C_7e^{kx}+\tilde C_8e^{-kx}]^2}}\\0&{{8k^2\tilde C_7\tilde C_8}\over{[\tilde C_7e^{kx}+\tilde C_8e^{-kx}]^2}}
\end{pmatrix}\nonumber\\
&+\begin{pmatrix}0&{{2(\tilde C_8+C_2\tilde C_7)}\over{[\tilde C_7e^{kx}+\tilde C_8e^{-kx}]^2}}+{{4C_2}\over{[e^{kx}+C_2e^{-kx}][\tilde C_7e^{kx}+\tilde C_8e^{-kx}]}}
\\0&0\end{pmatrix},\nonumber\\
\tilde\Psi_{1,0}^+(x)&={\bf C}^{-1}\Psi_{6,0}^+(x)=\begin{pmatrix}{{2kC_2}\over{e^{kx}+C_2e^{-kx}}}\\0\end{pmatrix},\nonumber\\
\tilde\Psi_{1,1}^+(x)&={\bf C}^{-1}\Psi_{6,1}^+(x)\!=\!\begin{pmatrix}{{[C_2x+kC_6][\tilde C_7e^{kx}-\tilde C_8e^{-kx}]}\over{[e^{kx}+C_2e^{-kx}][\tilde C_7e^{kx}+\tilde C_8e^{-kx}]}}\!-\!{{\tilde C_8+C_2\tilde C_7}\over{2k[\tilde C_7e^{kx}+\tilde C_8e^{-kx}]}}\!+\!{{kC_6}\over{e^{kx}+C_2e^{-kx}}}\\2k{{\tilde C_7\tilde C_8}\over{\tilde C_7e^{kx}+\tilde C_8e^{-kx}}}\end{pmatrix},\nonumber\\
\tilde\Psi_{1,1}^+(x)&\Big|_{C_2=C_4=0}={\bf C}^{-1}\Psi_{6,1}^+(x)\Big|_{C_2=C_4=0}=\begin{pmatrix}{{4k^2C_6\tilde C_7-\tilde C_8}\over{2k[\tilde C_7e^{kx}+\tilde C_8e^{-kx}]}}\\2k{{\tilde C_7\tilde C_8}\over{\tilde C_7e^{kx}+\tilde C_8e^{-kx}}}\end{pmatrix},\nonumber
\end{align}

\[{\bf C}=\begin{pmatrix}1&0\\C_3&-\alpha\end{pmatrix},\qquad{\bf C}^{-1}=\begin{pmatrix}1&0\\{C_3\over\alpha}&-{1\over\alpha}\end{pmatrix},\qquad\det{\bf C}=-\alpha,\qquad\alpha\in{\Bbb C},\,\,\,\alpha\ne0,\]
\[\tilde C_7=-{1\over\alpha}C_7,\qquad\tilde C_8=-{1\over\alpha}(C_8-C_3C_6),\]
where $\tilde W(x)$ is the Wronskian of $\tilde\Phi_1^-(x)$ and $\tilde\Phi_2^-(x)$.
It is evident that the representations and the intertwining (\ref{pres99}) transform trivially into the analogous formulae for the Hamiltonians ${\bf C}^{-1}H_+{\bf C}=H_+=-\partial^2$ and ${\bf C}^{-1}H_-{\bf C}$, for the matrix ${\bf C}^{-1}U_0(x){\bf C}$ and for the operators ${\bf C}^{-1}Q_1^+{\bf C}$ and ${\bf C}^{-1}Q_1^-{\bf C}$, that $\tilde\Psi_{1,0}^+(x)$ and $\tilde\Psi_{1,1}^+(x)$ for $|C_2|+|C_4|\ne0$ are vector-eigenfunction and associated vector-function of the first order (formal sometimes) respectively of the Hamiltonian ${\bf C}^{-1}H_-{\bf C}$ for the same eigenvalue $\lambda=-k^2$, that $\tilde\Psi_{1,1}^+(x)$ for $C_2=C_4=0$ is a vector-eigenfunction (formal sometimes) of the Hamiltonian ${\bf C}^{-1}H_-{\bf C}$ for the same eigenvalue $\lambda=-k^2$ and that $\tilde\Phi_1^-(x)$ and $\tilde\Phi_2^-(x)$ are transformation vector-functions corresponding to conversion of the Hamiltonian ${\bf C}^{-1}H_+{\bf C}=H_+$ to the Hamiltonian ${\bf C}^{-1}H_-{\bf C}$ with the help of the intertwining operator ${\bf C}^{-1}Q_1^-{\bf C}$.

\section{Conclusions}

In conclusion we itemize some problems which could be solved in future papers.
\renewcommand{\labelenumi}{\rm{(\theenumi)}}
\begin{enumerate}
\item To work out methods of spectral design for matrix Hamiltonians with the help of matrix intertwining operators of arbitrary order and, in particular, to find a criterion for transformation vector-functions that provides a desired changes for the spectrum of the corresponding final matrix Hamiltonian with respect to the spectrum of an initial matrix Hamiltonian. It is possible to try for this purpose to generalize Index Theorem and Lemma 4 of \cite{ancaso07,sokolov07np} to the matrix case.
\item To investigate (in)dependence of matrix differential intertwining operators in the way analogous to one of \cite{anso03} and, in particular, to define the notions of dependence and independence for these operators, to find a criterion of dependence for them and to solve the questions on maximal number of independent matrix differential intertwining operators and on a basis of such operators.
\item By analogy with \cite{anso03,anso09} to investigate in the matrix case properties of a minimal matrix differential hidden symmetry operator.

\item To investigate (ir)reducibility of matrix differential intertwining operators and, in particular, to classify irreducible and absolutely irreducible \cite{sokolov13} matrix differential intertwining operators in the way analogous to one of \cite{acdi95,anca04,samsonov99,anso06,sokolov07,sokolov10,ferneni00,tr89,dun98,khsu99,fermura03,fermiros02',fersahe03,samsonov06}.
\end{enumerate}

\section*{Acknowledgments}

The author is grateful to A.A. Andrianov for critical reading of this paper and valuable comments and to M.V. Ioffe for drawing attention to some papers on matrix models with supersymmetry. This work was supported by RFBR Grant 13-01-00136-a. The author acknowledges Saint-Petersburg State University for a research grant 11.38.660.2013.

\end{document}